\documentclass[%
 reprint,
superscriptaddress,
 amsmath,amssymb,
 prx,
noeprint,
longbibliography
]{revtex4-1}

\usepackage{bbold}
\usepackage{graphicx}
\usepackage{dcolumn}
\usepackage{bm}
\usepackage[colorlinks=true,linkcolor=blue,citecolor=blue,urlcolor=blue]{hyperref}

\usepackage{wasysym}

\newcommand{\sfg}[1]{\MakeLowercase{(#1)}}
\newcommand{\sfgp}[1]{\MakeLowercase{(#1)}}
\newcommand{\Methods}{\cite{Note1}}
\newcommand{\Supp}{\cite{Note1}}
\let\eqrefold\eqref
\renewcommand{\eqref}[1]{Eq.~\eqrefold{#1}}
\newcommand\bcirc{\raisebox{0.3mm}{$\Circle$}}

\newcommand{\beginsupplement}{%
        \setcounter{table}{0}
        \renewcommand{\thetable}{S\arabic{table}}%
        \setcounter{figure}{0}
        \renewcommand{\thefigure}{S\arabic{figure}}%
        \setcounter{equation}{0}
        \def\theequation{S\arabic{equation}}
     }

\begin{document}


\title{Optimal elasticity of biological networks}

\author{Henrik Ronellenfitsch}
\affiliation{Department of Mathematics, Massachusetts Institute of Technology, 77 Massachusetts Ave, Cambridge, MA 02139, USA}
\affiliation{Physics Department, Williams College, 33 Lab Campus Drive, Williamstown, MA 01267, USA}

\date{\today}

\begin{abstract}
Reinforced elastic sheets surround us in daily life, from concrete shell buildings
to biological structures such as the arthropod
exoskeleton or the venation network of dicotyledonous plant leaves.
Natural structures are often highly optimized through evolution and natural selection,
leading to the biologically and practically relevant problem of understanding
and applying the principles of their design.
Inspired by the hierarchically organized scaffolding networks found in
plant leaves, here we model networks of bending beams that
capture the discrete and
non-uniform nature of natural materials.
Using the principle of maximal rigidity under natural resource constraints,
we show that optimal discrete beam networks reproduce the structural
features of real leaf venation.
Thus, in addition to its ability to efficiently transport water and nutrients,
the venation network also optimizes leaf rigidity using
the same hierarchical reticulated network topology.
We study the phase space of optimal mechanical networks, providing concrete
guidelines for the construction of elastic structures.
We implement these natural design rules by fabricating efficient,
biologically inspired metamaterials.
\end{abstract}

\maketitle

Elastic sheets reinforced by beams are
pervasive in nature and engineering. From
concrete shell buildings~\cite{Melaragno2012} to aircraft fuselages~\cite{Niu1999},
reinforced shells have found numerous applications
due to their rigidity and efficient use of resources.
Evolution and natural selection have also produced structures
such as plant leaves, which need to remain flat to maximize photosynthesis~\cite{Niklas1994,Niklas1999,Roth-Nebelsick2001,Sack2013},
or dragonfly wings,
which combine light weight and rigidity to enable efficient flight~\cite{Sun2012}.
Uncovering the design rules behind biologically
optimized natural materials may not just impact engineering but also
illuminate their role in evolution.

Efficient design of thin shells is an active
research problem~\cite{Gil-Ureta2019,Sakai2020,Seranaj2018,Townsend2019,Bendsoe2003,Ramm1993,Hassani2013}, and
mechanical metamaterials have emerged as promising candidates for efficient, rigid and tunable structures~\cite{Bertoldi2017,Overvelde2017,Gross2019,Goodrich2015,Ronellenfitsch2018,Gurtner2014}.
Natural materials are often
characterized by a fractal-like hierarchical organization.
Specifically, the venation of plant leaves is known to
play a crucial role in the transport of water and nutrients~\cite{Katifori2018},
and in the structural rigidity of the
lamina~\cite{Sack2013,Roth-Nebelsick2001,Niklas1994}, so as to allow the plant
to maximize area for photosynthesis while
being compliant with the wind and
other forces~\cite{Ennos2005,Vogel2012}.
While much work has been done to characterize the venation networks of
dicotyledonous plants in terms of geometry~\cite{Wright2004,Blonder2011,Kull1994}, topology~\cite{Katifori2012,Mileyko2012,Ronellenfitsch2015b},
and optimal fluid transport~\cite{Katifori2010b,Corson2010flow,Hu2013,Ronellenfitsch2016,Ronellenfitsch2019,Gavrilchenko2019}, the mechanical
purpose, properties, and optimality of the venation network beyond the midrib~\cite{Niklas1994,Gilet2015,Niklas1999,Wei2012} have received less
attention~\cite{Sun2018}. Recent work points
towards the importance of mechanical traits~\cite{Blonder2020}.
\begin{figure*}
    \centering
    \includegraphics[width=.65\textwidth]{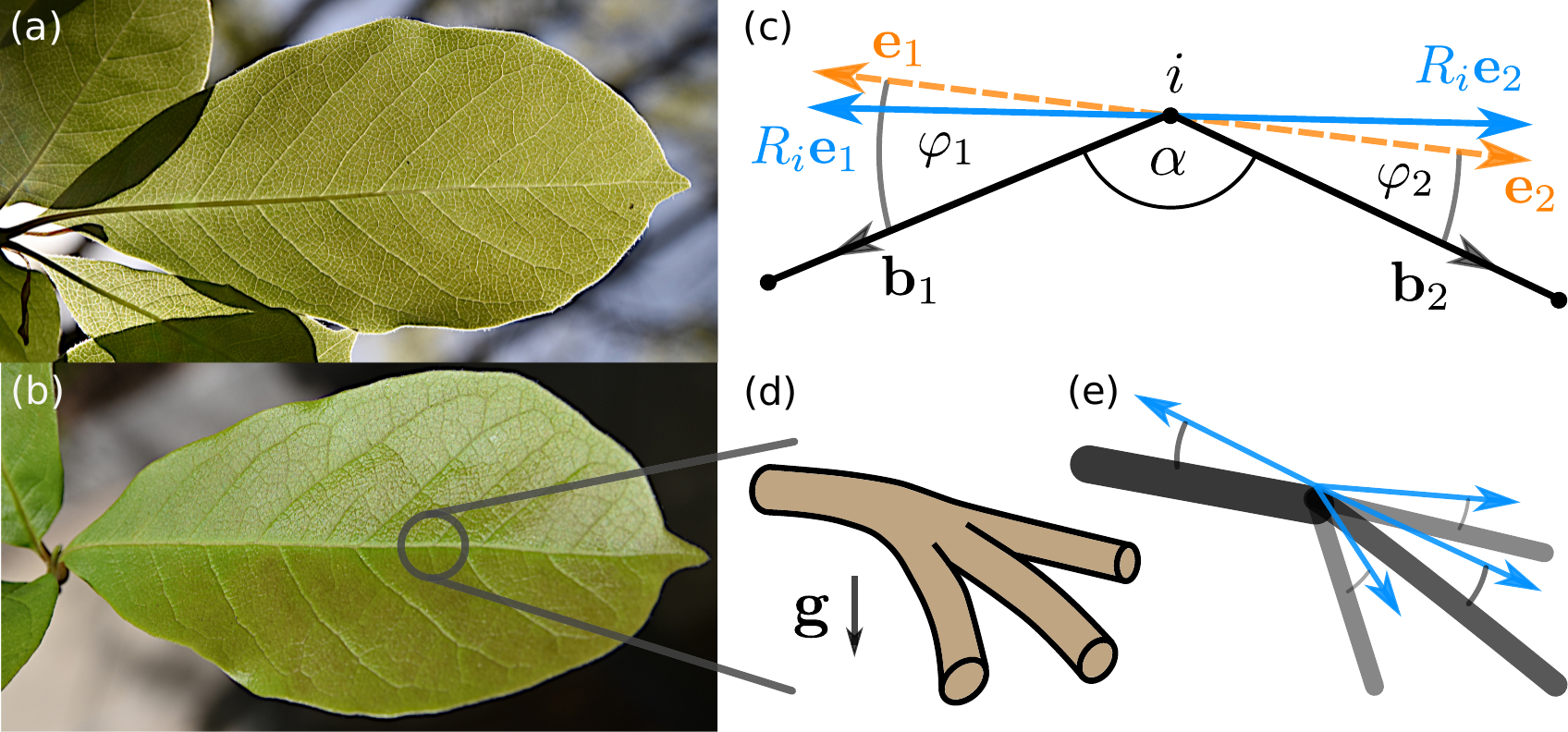}
    \caption{Leaf venation as a discrete beam network.
    \sfgp{A} Abaxial surface of a leaf of \emph{Magnolia sp.}, showing
    the hierarchically organized reticulate
    venation network keeping
    the lamina flat and rigid, and transporting water and nutrients. \sfgp{b} Adaxial surface of the same leaf, emphasizing the
    venation network embedded in the lamina.
    \sfgp{c} Discretized model of beam bending.
    Dashed orange arrows correspond to the local reference frame $\{\mathbf{e}_{1,2}\}$
    used to construct the elastic energy \eqref{eq:local-ref} with
    $\sin^2\varphi_{1,2} =
    \|\mathbf{e}_{1,2}\times\mathbf{b}_{1,2}\|^2$. The
    reference frame $\{R_i\,\mathbf{e}_{1,2}\}$
    compensating overall rigid rotations
    is shown in blue.
    \sfgp{d} Plant leaf venation subject to gravitational load
    $\mathbf{g}$ as prototypical example of a natural DBN.
    One large vein
    branches off into three smaller veins that all bend under the load.
    \sfgp{e} DBN model of the node from \sfg{d}. Each discrete beam
    joining at the node is depicted with its bending constant
    by line thickness and color. {Deviations from the local reference (blue) are penalized by \eqref{eq:dbn-with-rot}.}}
    \label{fig:beams-dbn}
\end{figure*}
Here, we ask to which extent leaves and similar
natural materials may be mechanically optimized, what rules their
natural design underlies, and how these rules can be applied.

To answer these questions,
we consider a model of discrete beam networks (DBNs) to
capture the properties of natural materials. Specifically,
DBNs model
bending beams with arbitrary stiffness that are joined to
form an elastic network.
We apply this generic model to the elasticity of leaf venation.
We numerically minimize the network's
compliance, maximizing overall rigidity under natural
loads~\cite{Bendsoe2003},
with a resource constraint to model the cost--efficiency trade-off that these networks are subject
to~\cite{Bohn2007,Durand2007,Savage2010,Blonder2011,Price2014b,Ronellenfitsch2019}.
We find that optimized mechanical DBNs exhibit similar structural features as real leaves: a
central midrib and hierarchically branching higher order veins connected
by anastomoses, in
close correspondence to vascular networks found
by optimizing for robust liquid transport~\cite{Murray1926,West1999,Katifori2010b,Katifori2018,Corson2010flow,Hu2013,Ronellenfitsch2016,Ronellenfitsch2019,Kirkegaard2020}.
Features of the leaf venation
such as the structure of interconnecting anastomoses and loops are thus
naturally explained by mechanical optimization.
We identify distinct topological phases as design rules
of optimal DBNs that
lead to substantially improved rigidity of the network, and use these rules
to design and manufacture efficient elastic metamaterials.

The theory of elastic sheets
connects curvature to an elastic energy~\cite{Safran1999,Helfrich1973}
and has been used with great success to model uniform
membranes and shells~\cite{Katifori2010,Couturier2013,Seung1988,Liang2009,Guckenberger2016,Gompper1996,Witten2007}.
Methods like topology optimization~\cite{Bendsoe2003} are tailored
for non-uniform continua, and progress has been made optimizing
reinforced elastic shells~\cite{Gil-Ureta2019,Sakai2020}.
We now consider a simple model of beam networks
that captures the discreteness and non-uniformity of
natural materials.
As an illustrative example, take a cylindrical
beam with bending energy~\cite{Audoly2010,Bergou2008},
\begin{align}
    V_b = \frac{\pi}{8} Y \ell r^4 \frac{1}{\mathcal{R}^2}
    \approx \frac{1}{2} \kappa \sin^2\alpha =
    \frac{1}{2} \kappa\, \|\mathbf{b}_1 \times \mathbf{b}_2\|^2,
    \label{eq:discrete-kirchhoff}
\end{align}
where $Y$ is the beam's Young's modulus, $r$ is its radius,
$\ell$ is its length, and $\mathcal{R}$ is its radius of
curvature.
The bending angle $\alpha$ was introduced by discretizing the
beam using the unit vectors $\mathbf{b}_{1,2}$
and approximating the curvature
\{Fig.~\ref{fig:beams-dbn}~\sfg{c}, Ref.~\footnote{See Supplemental Material [url] for a detailed discussion of the approximations, a derivation the elastic energy and the constrained optimization algorithm, the continuum limit, a discussion of mechanical constraints and optimization under self-loads, three-dimensional DBNs, and a comparison of optimal DBNs to real leaf networks using topological metrics, which includes
Refs.~\cite{DoCarmo1988,Lubensky2015,Bruyneel2005}}. The constants of proportionality
were combined into the bending constant $\kappa = \pi Y r^4/\ell$.
{It is possible to find an equivalent formulation
of \eqref{eq:discrete-kirchhoff} using two
elastically connected rigid beams}.

We introduce a {set of} unit vectors
$\{\mathbf{e}_1, \mathbf{e}_2\}$ at the midpoint, corresponding to the {reference} configuration
of the beams [Fig.~\ref{fig:beams-dbn}~\sfg{c}]. An elastic
energy {penalizing deviations from this reference} is
then,
\begin{align}
    V = \frac{1}{2} \kappa_b \| (R\,\mathbf{e}_1) \times \mathbf{b}_1 \|^2
    + \frac{1}{2} \kappa_b \| (R\,\mathbf{e}_2) \times \mathbf{b}_2 \|^2,
    \label{eq:local-ref}
\end{align}
{where $R$ is a rotation matrix.
This two-beam energy is equivalent to
\eqref{eq:discrete-kirchhoff} if $R$ is chosen to
compensate any overall rigid rotations, which can be found by minimizing $V$ over $R$ at fixed
$\mathbf{b}$~\Methods.}
\begin{figure*}[ht!]
    \centering
    \includegraphics[width=.75\textwidth]{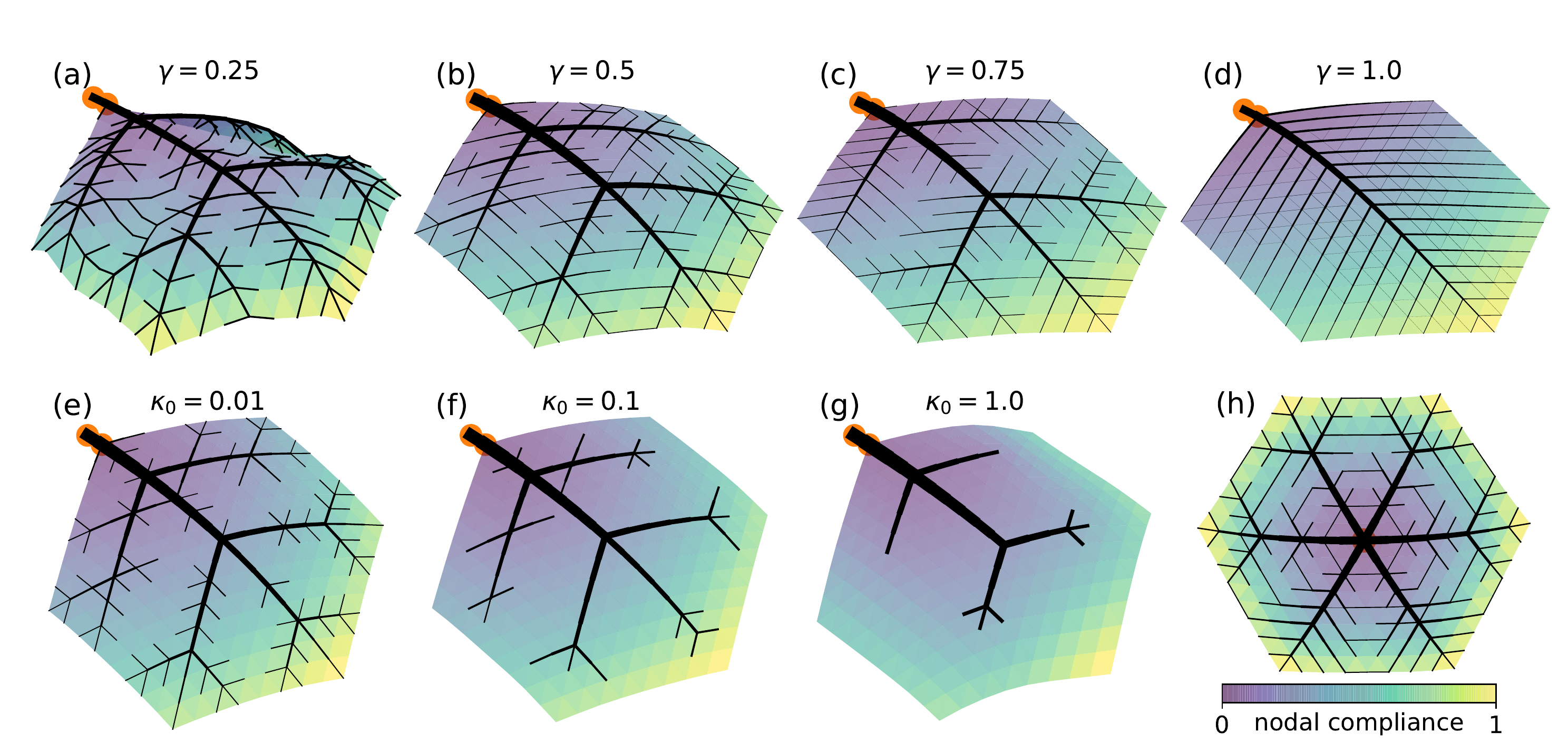}
    \caption{Compliance-optimized flat DBNs resemble real leaf venation.
    We optimized triangular DBNs with $N=217$ nodes and $E_\text{triang}=600$ edges.
    \sfgp{A--C} For $0< \gamma < 1$, optimal networks are sparse and show hierarchical
    organization and anastomosing reticulation. \sfgp{D} At the transition $\gamma=1$ the network
    becomes highly reticulate and less hierarchically organized.
    The networks \sfg{A--D} were subject to a uniform downward load,
    the petiole was modeled as one additional node the position of which
    was fixed, and overall twists of the petiole were removed.
    The lamina stiffness was $\kappa_0 = 10^{-6}$.
    \sfgp{E--G} Optimal networks reduce to just the main veins
    as the lamina stiffness $\kappa_0$ is increased.
    \sfgp{H} An optimal network with the petiole at the center and subject to
    a uniform upward load.
    The cost parameter in \sfg{E--H} was $\gamma=1/2$, and
    the lamina stiffness in \sfg{H}
    was $\kappa_0 = 10^{-6}$.
    Fixed nodes are shown as red dots, each triangle is colored by the average
    {nodal compliance $\mathbf{f}_i^\top \mathbf{u}_i$} of the adjacent nodes normalized by the maximum, and the line thicknesses are
    proportional to $\kappa_b^{\gamma/2}$.}
    \label{fig:leaves}
\end{figure*}
{Equation~(\ref{eq:local-ref}) then suggests that the
elastic energy of an arbitrary number
of beams elastically connected at a node $i$
[Fig.~\ref{fig:beams-dbn}~(e)] can be written as,}
\begin{align}
    V_i = \frac{1}{2} \sum_{b\in B_i} \kappa_b \| (R_i\, \mathbf{e}_b)
                \times \mathbf{b} \|^2,
                \label{eq:dbn-with-rot}
\end{align}
where the sum runs over the set $B_i$ of edges joining at node $i$,
$\kappa_b$ is the bending constant of edge $b$, and $\mathbf{b}$ is
the unit vector pointing from node $i$ to node $j$ along the edge $b=(ij)$. The node's equilibrium
configuration is given by the local reference frame $\{ \mathbf{e}_b \}_{b\in B_i}$
and
$R_i$ {compensates overall rigid rotations}.
We now linearize \eqref{eq:dbn-with-rot} by expanding both
$R_i$ and $\mathbf{b}$ and minimizing over $R_i$~\Methods. We find for a network consisting of $N$ nodes,
\begin{align}
    V &= \frac{1}{2} \mathbf{u}^\top \left(
    H_\text{eq} - H_\text{or} \right) \mathbf{u} =
    \frac{1}{2} \mathbf{u}^\top H \mathbf{u}, \label{eq:dbn-bending-final}
\end{align}
where $\mathbf{u}$ is the $3N$-dimensional vector of
nodal displacements from equilibrium.
The term $(1/2)\,\mathbf{u}^\top H_\text{eq}\mathbf{u}$
is the elastic energy with respect to the {fixed}
equilibrium frame $\{\mathbf{e}_b\}$, while $(1/2)\,\mathbf{u}^\top H_\text{or}\mathbf{u}$
corrects for {overall rotations}~\Methods.
Given {any static} loads $\mathbf{f}$
on the network, the displacements satisfy $H \mathbf{u} = \mathbf{f}$.
At each node, this force balance can be expressed as
$\mathbf{f}_i = \sum_{j} (\mathbf{F}_{ij} - \mathbf{F}_{ji})$,
where $\mathbf{F}_{ij}$ is the force on node $i$ due to the connection
to node $j$, and $\mathbf{f}_i$
is the load on node $i$~\Methods.

While our model applies
to generic elastic networks, we now specialize to leaf-like
structures.
We consider planar DBNs
described by \eqref{eq:dbn-bending-final} {and embedded in an inextensible lamina.}
Inextensibility {of both beam network and lamina}
is implemented to linear order by allowing only nodal
displacements $\mathbf{u}$ that satisfy
$\mathbf{e}_b^\top (\mathbf{u}_j-\mathbf{u}_i)=0$ for all edges $b$~{\cite{Seung1988,Witten2007,Note1}.}

Leaves must remain flat and rigid to present a maximal area to sunlight
for photosynthesis. Thus, we expect the reinforced scaffolding
network to be optimized
under the influence of gravitational or wind load.
Maximum rigidity of a mechanical system under loads
$\mathbf{f}$ leading to displacements $\mathbf{u}$ corresponds to
minimum compliance
$c=\mathbf{f}^\top \mathbf{u} = \sum_i
\mathbf{f}_i ^\top \mathbf{u}_i$~\cite{Bendsoe2003},
{where $\mathbf{f}_i$ is the load on node $i$ and $\mathbf{u}_i$ is its displacement.}
In the following, we minimize the
compliance over the set of bending constants $\kappa_b$ of the network.
Biological networks are constrained by the
amount of resources available, and by the requirement to distribute
them efficiently. Following Refs.~\cite{Bohn2007,Durand2007,Katifori2010b,Corson2010flow,Hu2013,Ronellenfitsch2016},
we incorporate this by introducing the constraint $\sum_{b} \kappa_b^\gamma = K$, where the parameter $\gamma$ models the material cost of each beam and
$K$ is the overall cost.
A natural material constraint is the total mass of the network,
which for beams following \eqref{eq:discrete-kirchhoff} corresponds to
$\gamma=1/2$. More generally, $0<\gamma <1$ leads to an economy
of scale promoting sparse networks~\cite{Chartrand2007}.
We now focus on this biologically relevant regime.

\begin{figure*}
    \centering
    \includegraphics[width=.85\textwidth]{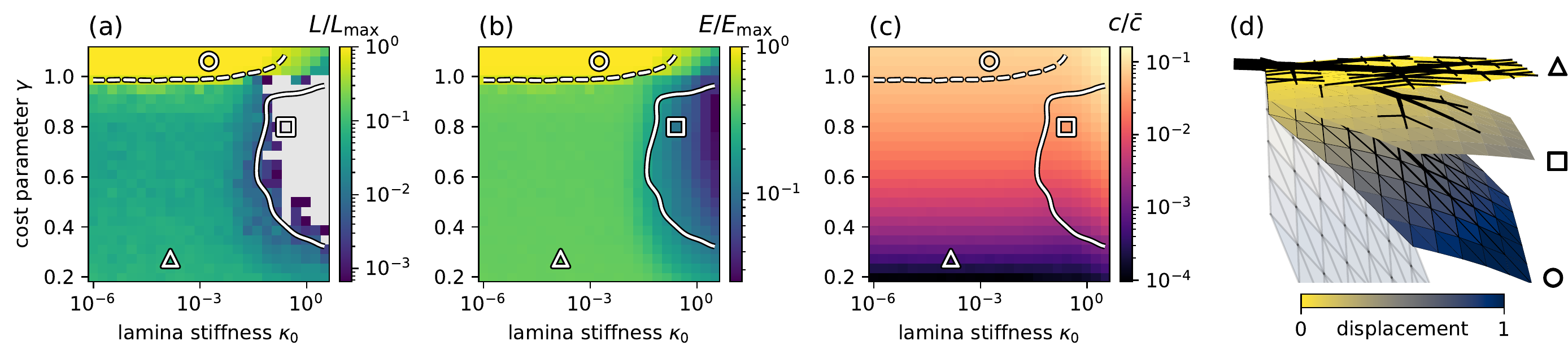}
    \caption{Topological transition and phase space of optimal
    DBNs with leaf boundary conditions. Each pixel in the
    $25\times 25$ images \sfg{A--C} corresponds
    to a mean over 10 annealed triangular networks with $N=92$ nodes and
    $E_\text{max}=241$ edges. \sfgp{a} Network topology is encoded in the loop density
    $L/L_\text{max}$, where $L$ is the number of loops and
    $L_\text{max}=150$ is the maximum number of loops in the
    triangular grid. Grey pixels correspond to $L=0$.
    The dashed and solid lines approximately mark the
    transitions to maximally loopy and tree topologies,
    respectively.
    \sfgp{B} Network structure as measured by the number of nonzero bending constant
    edges $E$ normalized by the maximum number $E_\text{max}$ of edges in the
    triangular grid.
    \sfgp{C} The compliance $c$ of the optimized
    networks, normalized by the compliance $\bar{c}$ of
    a uniform network with identical cost $K$.
    The results in \sfg{A--C} remain qualitatively valid for larger networks as well~\Supp.
    \sfgp{D} Optimal networks $\bigtriangleup$, $\square$, \bcirc, and a
    uniform network
    shown with their relative displacements under the same load.
    The optimal networks are also marked in panels \sfg{A--C}.
    Displacements are measured relative
    to the tip of network \bcirc.}
    \label{fig:topology}
\end{figure*}

The optimal $\kappa_b$ are encoded in a scaling relation with the nodal forces~\Supp,
\begin{align}
    \kappa_{b} \sim \left(\ell_{b}^2\,(\|\mathbf{F}_{ij}\|^2 + \|\mathbf{F}_{ji}\|^2)\right)^{\frac{1}{1+\gamma}},
    \label{eq:scaling}
\end{align}
where the edge $b$ connects nodes $i$ and $j$.
To avoid local minima due to the non-convex constraint,
we employ a numerical optimization algorithm based on
simulated annealing~\Methods.
In the following, we start from a triangular grid in the $x$--$y$ plane representing
the leaf lamina, which is attached to
a petiole with fixed position and orientation
\{Fig.~\ref{fig:leaves}, Ref.~\Supp\}.
The entire leaf is subject to uniform load in the negative
$z$ direction [Fig.~\ref{fig:beams-dbn}~\sfg{d,e}],
{such that the compliance is now proportional to the average displacement. This is a reasonable approximation
given typical leaf mass composition~\cite{John2017}.
Including vein self-loads in this regime does not lead
to markedly different optimal networks~\cite{Note1}.}
Because the leaf lamina itself is rigid, we set the
bending constants to $\kappa_0 + \kappa_b$, where
$\kappa_0$ is the lamina stiffness and the $\kappa_b$ are the
bending constants of the network that we minimize over.
The inextensibility constraint is enforced on all edges of the triangular grid irrespective of their bending rigidity, such that the lamina is always inextensible to linear order.
The cost $K$ is fixed to the number of edges in the triangular grid, setting the
scale for the $\kappa_b$.
We first specialize to the regime $\kappa_0 = 10^{-6} \ll \kappa_b$ where the elastic properties
are dominated by the venation network. Here, optimized DBNs are rigid and flat, decreasing
the compliance by a factor of $\sim 100$ compared
to uniform networks [Fig.~\ref{fig:topology}~\sfg{c}].
Their structure {exhibits the basic features of} dicotyledonous leaf venation [Fig.~\ref{fig:leaves}~\sfg{A--E}],
including a hierarchical midrib and branching and anastomosing higher order veins. {This is also reflected in
quantitative topological measures when comparing to real
leaf networks~\Supp.}
Mechanically optimized DBNs are structurally similar to
distribution networks optimizing robust fluid
transport~\cite{Katifori2010b,Corson2010flow,Ronellenfitsch2016,Hu2013,Ronellenfitsch2019}.
This is due to a connection between hydraulic and elastic leaf {network models,
both of which can be seen as
conservation laws (of fluid or force)
with a single source and many sinks.}
Under the inextensibility constraint,
$(1/2)\mathbf{u}^\top H_\mathrm{eq} \mathbf{u} = \sum_{i,j} (\kappa_b/\ell_b^2)\, (u_{z,j} - u_{z,i})^2$, where
$u_{z,i}$ are the $z$ components of the displacements~\Supp.
Formally identifying $\kappa_b/\ell_b^2$ with the hydraulic conductivity and
the perpendicular displacements $u_{z,i}$ with the potential,
this part of the compliance has the same form as the power dissipation
minimized for flow networks {and encodes only
the weighted network topology}.
Optimal flow networks are known to correspond to topological trees~\cite{Banavar2000}, even though the
global optimum may not be hierarchical~\cite{Yan2018}.
Thus, the {geometric} term
$\mathbf{u}^\top H_\mathrm{or} \mathbf{u}$ is responsible
for departure from the tree-optima and induces redundant
connections in mechanical networks~\Supp.
This intrinsic elastic mechanism stands in contrast to
flow networks where only explicitly modeling additional
effects such as resistance to fluctuations or damage
can induce loops~\cite{Katifori2010,Hu2013,Corson2010flow}.

When $\gamma > 1$, the optimization problem becomes convex, and a single
global minimum exists, containing a midrib but
otherwise appearing featureless [Fig.~\ref{fig:leaves}~\sfg{D}].
The generic properties of optimal DBNs remain valid for other boundary conditions
as well~[Fig.~\ref{fig:leaves}~\sfg{H}].

We now proceed to study the topological transition
from non-reticulate to reticulate optimal networks. The topology of planar
networks is quantified by the number of loops $L = E - N + 1$, as obtained
from Euler's formula. Optimal DBNs
exhibit three basic topological phases [Fig.~\ref{fig:topology}~\sfg{A}].
In the convex regime where $\gamma>1$ and
the lamina stiffness $\kappa_0 \lesssim 10^{-2}$, the optimal networks
corresponding to the single global minimum are maximally loopy.
As $\gamma$ is decreased below $1$, most loops are
lost and the optimal
networks feature a small number of loops that is approximately constant
over a wide range of parameter values. Increasing the lamina stiffness
beyond $\kappa_0 \approx 10^{-2}$ leads to a gradual crossover into a loop-less regime, where only main and secondary veins are reinforced [Fig.~\ref{fig:leaves}~\sfg{E--G}]. These transitions are mirrored in the number of nonzero bending constant
edges $E$ in the network, with the difference that $E$ gradually decreases
as $\kappa_0$ is increased instead of dropping to zero [Fig.~\ref{fig:topology}~\sfg{B}].
Surprisingly, the optimal compliance does not vary strongly
with the optimal network topology [Fig.~\ref{fig:topology}~\sfg{C, D}].
Instead, the optimal compliance is largely independent of the
lamina stiffness $\kappa_0$ and varies strongly only with the cost parameter
$\gamma$. Since $\gamma$ is expected to be fixed by
geometry, this suggests that generically, it pays
to invest in an optimized mechanical network, even if this means only
reinforcing the main vein. Even then, the improvement in compliance is significant [Fig.~\ref{fig:topology}~\sfg{C}].

\begin{figure}
    \centering
    \includegraphics[width=.49\textwidth]{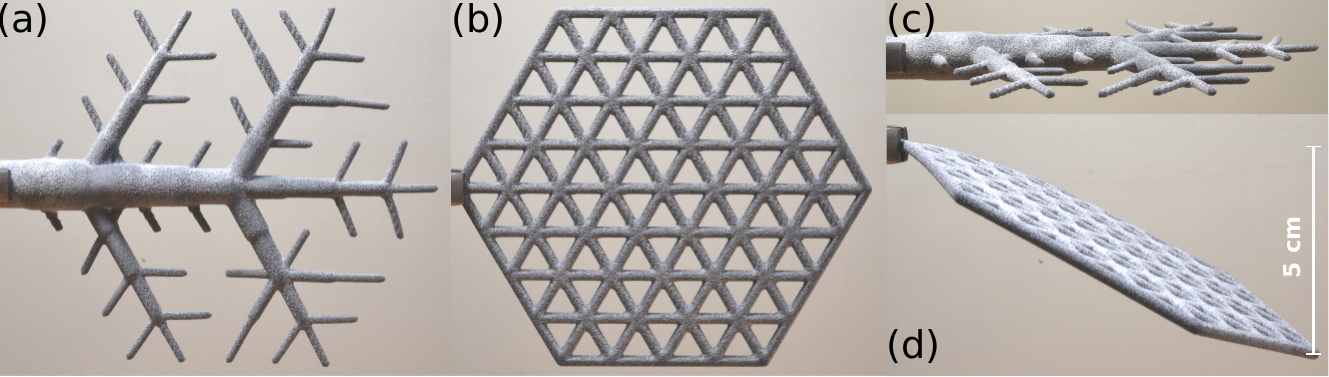}
    \caption{Biologically inspired metamaterials for flatness and rigidity.
    \sfgp{A} Additively manufactured metamaterial based on an optimized
    DBN topology with $\gamma=1/2$. The vertical size is $11\,\mathrm{cm}$,
    the material is thermoplastic polyurethane~\Methods. \sfgp{B} Metamaterial based
    on a uniform DBN topology with equal size and total volume.
    Beam radii in \sfg{A, B} are proportional to $\kappa_b^{1/4}$, and $\kappa_0=0$.
    \sfgp{c, d} Side view of the same networks
    clamped at the petiole. The optimized network \sfg{c}
    remained flat. The uniform network \sfg{d} showed a tip displacement
    of approximately $5\,\mathrm{cm}$.
    }
    \label{fig:metamaterials}
\end{figure}

The natural design principles of leaf venation can be applied to the design
of efficient rigid metamaterials.
We additively manufactured
networks of connected cylindrical
beams based on optimized and uniform DBN topologies
with equal material volume \{Fig.~\ref{fig:metamaterials}~\sfg{A, B}, Ref.~\Methods\}. The improvement in rigidity in the optimized manufactured network is
significant, with no bending or tip displacement discernible
[Fig.~\ref{fig:metamaterials}~\sfg{C}].
This is compared to the uniform network, which bends visibly
[Fig.~\ref{fig:metamaterials}~\sfg{D}].
This suggests that biologically inspired elastic networks
may provide design principles for discrete metamaterials.

In summary, we considered a model of discrete beam networks that is able to naturally
represent non-uniform reinforcing scaffoldings of elastic sheets
and networks, and applied it to leaf venation.
We showed that optimal DBNs
minimizing mechanical compliance under a cost constraint
resemble real leaves, including a hierarchical backbone,
anastomoses, and loops between the veins.
Using the principles learned from nature, we designed and manufactured
elastic metamaterials.

Our results may have implications for the biology of leaves
and other natural materials with a combined mechanical and
hydraulic function such as dragonfly wings~\cite{Sun2012}.
The relevance of fluid flow optimization for leaf venation is well-known when rationalizing loops as an evolutionary adaptation to damage or
fluctuations~\cite{Katifori2018,Ronellenfitsch2019}.
At the same time, the reduction in compliance of optimized over uniform DBNs is
highly significant.
Thus, maximizing stiffness could result in an evolutionary advantage.
Leaves are therefore in the extraordinary position to optimize
two highly disparate requirements, mechanical rigidity and robust
fluid transport, using the same hierarchically organized, reticulate venation network
architecture.
Our results may also offer a connection between the differing approaches
modeling leaf vascular development as adaptive
mechanisms relying on either flow~\cite{Runions2014,Dimitrov2006,Ronellenfitsch2016} or mechanical~\cite{Corson2010mechanics,Corson2009,Laguna2008,Bar-Sinai2016,Couder2002} cues.
More generally, our work paves the way for detailed study of optimized mechanical networks
in other biological systems such as actin-myosin networks~\cite{Mizuno2007},
active mechanics~\cite{Ronceray2016,Noll2017}, allosteric materials~\cite{Rocks2017}, or
network control~\cite{Kim2019}.

\acknowledgments{The author wishes to thank
Ellen A. Donnelly for helpful
discussions and the MIT Department of Mathematics for support.}

\bibliography{references}

\providecommand{\noopsort}[1]{}\providecommand{\singleletter}[1]{#1}%
\begin{thebibliography}{77}%
\makeatletter
\providecommand \@ifxundefined [1]{%
 \@ifx{#1\undefined}
}%
\providecommand \@ifnum [1]{%
 \ifnum #1\expandafter \@firstoftwo
 \else \expandafter \@secondoftwo
 \fi
}%
\providecommand \@ifx [1]{%
 \ifx #1\expandafter \@firstoftwo
 \else \expandafter \@secondoftwo
 \fi
}%
\providecommand \natexlab [1]{#1}%
\providecommand \enquote  [1]{``#1''}%
\providecommand \bibnamefont  [1]{#1}%
\providecommand \bibfnamefont [1]{#1}%
\providecommand \citenamefont [1]{#1}%
\providecommand \href@noop [0]{\@secondoftwo}%
\providecommand \href [0]{\begingroup \@sanitize@url \@href}%
\providecommand \@href[1]{\@@startlink{#1}\@@href}%
\providecommand \@@href[1]{\endgroup#1\@@endlink}%
\providecommand \@sanitize@url [0]{\catcode `\\12\catcode `\$12\catcode
  `\&12\catcode `\#12\catcode `\^12\catcode `\_12\catcode `\%12\relax}%
\providecommand \@@startlink[1]{}%
\providecommand \@@endlink[0]{}%
\providecommand \url  [0]{\begingroup\@sanitize@url \@url }%
\providecommand \@url [1]{\endgroup\@href {#1}{\urlprefix }}%
\providecommand \urlprefix  [0]{URL }%
\providecommand \Eprint [0]{\href }%
\providecommand \doibase [0]{http://dx.doi.org/}%
\providecommand \selectlanguage [0]{\@gobble}%
\providecommand \bibinfo  [0]{\@secondoftwo}%
\providecommand \bibfield  [0]{\@secondoftwo}%
\providecommand \translation [1]{[#1]}%
\providecommand \BibitemOpen [0]{}%
\providecommand \bibitemStop [0]{}%
\providecommand \bibitemNoStop [0]{.\EOS\space}%
\providecommand \EOS [0]{\spacefactor3000\relax}%
\providecommand \BibitemShut  [1]{\csname bibitem#1\endcsname}%
\let\auto@bib@innerbib\@empty
\bibitem [{\citenamefont {Melaragno}(2012)}]{Melaragno2012}%
  \BibitemOpen
  \bibfield  {author} {\bibinfo {author} {\bibfnamefont {Michele}\ \bibnamefont
  {Melaragno}},\ }\href {\doibase 10.1007/978-1-4757-0223-1} {\emph {\bibinfo
  {title} {An Introduction to Shell Structures: The Art and Science of
  Vaulting}}}\ (\bibinfo  {publisher} {Springer},\ \bibinfo {address} {Boston,
  MA},\ \bibinfo {year} {2012})\BibitemShut {NoStop}%
\bibitem [{\citenamefont {Niu}\ and\ \citenamefont {Niu}(1999)}]{Niu1999}%
  \BibitemOpen
  \bibfield  {author} {\bibinfo {author} {\bibfnamefont {C.}~\bibnamefont
  {Niu}}\ and\ \bibinfo {author} {\bibfnamefont {M.C.Y.}\ \bibnamefont {Niu}},\
  }\href {https://books.google.com/books?id=yT15SwAACAAJ} {\emph {\bibinfo
  {title} {Airframe Structural Design: Practical Design Information and Data on
  Aircraft Structures}}},\ Airframe book series\ (\bibinfo  {publisher} {Adaso
  Adastra Engineering Center},\ \bibinfo {year} {1999})\BibitemShut {NoStop}%
\bibitem [{\citenamefont {Niklas}(1992)}]{Niklas1994}%
  \BibitemOpen
  \bibfield  {author} {\bibinfo {author} {\bibfnamefont {Karl~J.}\ \bibnamefont
  {Niklas}},\ }\href {\doibase 10.2307/2807461} {\emph {\bibinfo {title}
  {{Plant Biomechanics: An Engineering Approach to Plant Form and Function}}}}\
  (\bibinfo  {publisher} {The University of Chicago Press},\ \bibinfo {address}
  {Chicago, London},\ \bibinfo {year} {1992})\BibitemShut {NoStop}%
\bibitem [{\citenamefont {Niklas}(1999)}]{Niklas1999}%
  \BibitemOpen
  \bibfield  {author} {\bibinfo {author} {\bibfnamefont {Karl~J.}\ \bibnamefont
  {Niklas}},\ }\bibfield  {title} {\enquote {\bibinfo {title} {{A mechanical
  perspective on foliage leaf form and function}},}\ }\href {\doibase
  10.1046/j.1469-8137.1999.00441.x} {\bibfield  {journal} {\bibinfo  {journal}
  {New Phytologist}\ }\textbf {\bibinfo {volume} {143}},\ \bibinfo {pages}
  {19--31} (\bibinfo {year} {1999})}\BibitemShut {NoStop}%
\bibitem [{\citenamefont {Roth-Nebelsick}\ \emph {et~al.}(2001)\citenamefont
  {Roth-Nebelsick}, \citenamefont {Uhl}, \citenamefont {Mosbrugger},\ and\
  \citenamefont {Kerp}}]{Roth-Nebelsick2001}%
  \BibitemOpen
  \bibfield  {author} {\bibinfo {author} {\bibfnamefont {Anita}\ \bibnamefont
  {Roth-Nebelsick}}, \bibinfo {author} {\bibfnamefont {Dieter}\ \bibnamefont
  {Uhl}}, \bibinfo {author} {\bibfnamefont {Volker}\ \bibnamefont
  {Mosbrugger}}, \ and\ \bibinfo {author} {\bibfnamefont {Hans}\ \bibnamefont
  {Kerp}},\ }\bibfield  {title} {\enquote {\bibinfo {title} {{Evolution and
  Function of Leaf Venation Architecture: A Review}},}\ }\href {\doibase
  10.1006/anbo.2001.1391} {\bibfield  {journal} {\bibinfo  {journal} {Annals of
  Botany}\ }\textbf {\bibinfo {volume} {87}},\ \bibinfo {pages} {553--566}
  (\bibinfo {year} {2001})}\BibitemShut {NoStop}%
\bibitem [{\citenamefont {Sack}\ and\ \citenamefont
  {Scoffoni}(2013)}]{Sack2013}%
  \BibitemOpen
  \bibfield  {author} {\bibinfo {author} {\bibfnamefont {Lawren}\ \bibnamefont
  {Sack}}\ and\ \bibinfo {author} {\bibfnamefont {Christine}\ \bibnamefont
  {Scoffoni}},\ }\bibfield  {title} {\enquote {\bibinfo {title} {{Leaf
  venation: structure, function, development, evolution, ecology and
  applications in the past, present and future}},}\ }\href {\doibase
  10.1111/nph.12253} {\bibfield  {journal} {\bibinfo  {journal} {New
  Phytologist}\ }\textbf {\bibinfo {volume} {198}},\ \bibinfo {pages}
  {983--1000} (\bibinfo {year} {2013})}\BibitemShut {NoStop}%
\bibitem [{\citenamefont {Sun}\ and\ \citenamefont {Bhushan}(2012)}]{Sun2012}%
  \BibitemOpen
  \bibfield  {author} {\bibinfo {author} {\bibfnamefont {Jiyu}\ \bibnamefont
  {Sun}}\ and\ \bibinfo {author} {\bibfnamefont {Bharat}\ \bibnamefont
  {Bhushan}},\ }\bibfield  {title} {\enquote {\bibinfo {title} {{The structure
  and mechanical properties of dragonfly wings and their role on
  flyability}},}\ }\href {\doibase 10.1016/j.crme.2011.11.003} {\bibfield
  {journal} {\bibinfo  {journal} {Comptes Rendus M{\'{e}}canique}\ }\textbf
  {\bibinfo {volume} {340}},\ \bibinfo {pages} {3--17} (\bibinfo {year}
  {2012})}\BibitemShut {NoStop}%
\bibitem [{\citenamefont {Gil-Ureta}\ \emph {et~al.}(2019)\citenamefont
  {Gil-Ureta}, \citenamefont {Pietroni},\ and\ \citenamefont
  {Zorin}}]{Gil-Ureta2019}%
  \BibitemOpen
  \bibfield  {author} {\bibinfo {author} {\bibfnamefont {Francisca}\
  \bibnamefont {Gil-Ureta}}, \bibinfo {author} {\bibfnamefont {Nico}\
  \bibnamefont {Pietroni}}, \ and\ \bibinfo {author} {\bibfnamefont {Denis}\
  \bibnamefont {Zorin}},\ }\href {http://arxiv.org/abs/1904.12240} {\enquote
  {\bibinfo {title} {{Structurally optimized shells}},}\ } (\bibinfo {year}
  {2019})\BibitemShut {NoStop}%
\bibitem [{\citenamefont {Sakai}\ \emph {et~al.}(2020)\citenamefont {Sakai},
  \citenamefont {Ohsaki},\ and\ \citenamefont {Adriaenssens}}]{Sakai2020}%
  \BibitemOpen
  \bibfield  {author} {\bibinfo {author} {\bibfnamefont {Yusuke}\ \bibnamefont
  {Sakai}}, \bibinfo {author} {\bibfnamefont {Makoto}\ \bibnamefont {Ohsaki}},
  \ and\ \bibinfo {author} {\bibfnamefont {Sigrid}\ \bibnamefont
  {Adriaenssens}},\ }\bibfield  {title} {\enquote {\bibinfo {title} {{A
  3-dimensional elastic beam model for form-finding of bending-active
  gridshells}},}\ }\href {\doibase 10.1016/j.ijsolstr.2020.02.034} {\bibfield
  {journal} {\bibinfo  {journal} {International Journal of Solids and
  Structures}\ }\textbf {\bibinfo {volume} {193-194}},\ \bibinfo {pages}
  {328--337} (\bibinfo {year} {2020})}\BibitemShut {NoStop}%
\bibitem [{\citenamefont {Seranaj}\ \emph {et~al.}(2018)\citenamefont
  {Seranaj}, \citenamefont {Elezi},\ and\ \citenamefont
  {Seranaj}}]{Seranaj2018}%
  \BibitemOpen
  \bibfield  {author} {\bibinfo {author} {\bibfnamefont {Agim}\ \bibnamefont
  {Seranaj}}, \bibinfo {author} {\bibfnamefont {Erald}\ \bibnamefont {Elezi}},
  \ and\ \bibinfo {author} {\bibfnamefont {Altin}\ \bibnamefont {Seranaj}},\
  }\bibfield  {title} {\enquote {\bibinfo {title} {{Structural optimization of
  reinforced concrete spatial structures with different structural openings and
  forms}},}\ }\href {\doibase 10.17515/resm2016.79st0726} {\bibfield  {journal}
  {\bibinfo  {journal} {Research on Engineering Structures and Materials}\
  }\textbf {\bibinfo {volume} {4}},\ \bibinfo {pages} {79--89} (\bibinfo {year}
  {2018})}\BibitemShut {NoStop}%
\bibitem [{\citenamefont {Townsend}\ and\ \citenamefont
  {Kim}(2019)}]{Townsend2019}%
  \BibitemOpen
  \bibfield  {author} {\bibinfo {author} {\bibfnamefont {Scott}\ \bibnamefont
  {Townsend}}\ and\ \bibinfo {author} {\bibfnamefont {H.~Alicia}\ \bibnamefont
  {Kim}},\ }\bibfield  {title} {\enquote {\bibinfo {title} {{A level set
  topology optimization method for the buckling of shell structures}},}\ }\href
  {\doibase 10.1007/s00158-019-02374-9} {\bibfield  {journal} {\bibinfo
  {journal} {Structural and Multidisciplinary Optimization}\ }\textbf {\bibinfo
  {volume} {60}},\ \bibinfo {pages} {1783--1800} (\bibinfo {year}
  {2019})}\BibitemShut {NoStop}%
\bibitem [{\citenamefont {Bends{\o}e}\ and\ \citenamefont
  {Sigmund}(2003)}]{Bendsoe2003}%
  \BibitemOpen
  \bibfield  {author} {\bibinfo {author} {\bibfnamefont {Martin~Philip}\
  \bibnamefont {Bends{\o}e}}\ and\ \bibinfo {author} {\bibfnamefont
  {O.}~\bibnamefont {Sigmund}},\ }\href
  {http://www.amazon.fr/Topology-Optimization-Theory-Methods-Applications/dp/3540429921}
  {\emph {\bibinfo {title} {{Topology optimization: theory, methods, and
  applications}}}},\ \bibinfo {edition} {2nd}\ ed.\ (\bibinfo  {publisher}
  {Springer},\ \bibinfo {address} {Berlin, Heidelberg, New York},\ \bibinfo
  {year} {2003})\BibitemShut {NoStop}%
\bibitem [{\citenamefont {Ramm}\ \emph {et~al.}(1993)\citenamefont {Ramm},
  \citenamefont {Bletzinger},\ and\ \citenamefont {Reitinger}}]{Ramm1993}%
  \BibitemOpen
  \bibfield  {author} {\bibinfo {author} {\bibfnamefont {Ekkehard}\
  \bibnamefont {Ramm}}, \bibinfo {author} {\bibfnamefont {Kai-Uwe}\
  \bibnamefont {Bletzinger}}, \ and\ \bibinfo {author} {\bibfnamefont {Reiner}\
  \bibnamefont {Reitinger}},\ }\bibfield  {title} {\enquote {\bibinfo {title}
  {{Shape optimization of shell structures}},}\ }\href {\doibase
  10.1080/12506559.1993.10511083} {\bibfield  {journal} {\bibinfo  {journal}
  {Revue Europ{\'{e}}enne des {\'{E}}l{\'{e}}ments Finis}\ }\textbf {\bibinfo
  {volume} {2}},\ \bibinfo {pages} {377--398} (\bibinfo {year}
  {1993})}\BibitemShut {NoStop}%
\bibitem [{\citenamefont {Hassani}\ \emph {et~al.}(2013)\citenamefont
  {Hassani}, \citenamefont {Tavakkoli},\ and\ \citenamefont
  {Ghasemnejad}}]{Hassani2013}%
  \BibitemOpen
  \bibfield  {author} {\bibinfo {author} {\bibfnamefont {Behrooz}\ \bibnamefont
  {Hassani}}, \bibinfo {author} {\bibfnamefont {Seyed~Mehdi}\ \bibnamefont
  {Tavakkoli}}, \ and\ \bibinfo {author} {\bibfnamefont {Hossein}\ \bibnamefont
  {Ghasemnejad}},\ }\bibfield  {title} {\enquote {\bibinfo {title}
  {{Simultaneous shape and topology optimization of shell structures}},}\
  }\href {\doibase 10.1007/s00158-013-0894-9} {\bibfield  {journal} {\bibinfo
  {journal} {Structural and Multidisciplinary Optimization}\ }\textbf {\bibinfo
  {volume} {48}},\ \bibinfo {pages} {221--233} (\bibinfo {year}
  {2013})}\BibitemShut {NoStop}%
\bibitem [{\citenamefont {Bertoldi}\ \emph {et~al.}(2017)\citenamefont
  {Bertoldi}, \citenamefont {Vitelli}, \citenamefont {Christensen},\ and\
  \citenamefont {van Hecke}}]{Bertoldi2017}%
  \BibitemOpen
  \bibfield  {author} {\bibinfo {author} {\bibfnamefont {Katia}\ \bibnamefont
  {Bertoldi}}, \bibinfo {author} {\bibfnamefont {Vincenzo}\ \bibnamefont
  {Vitelli}}, \bibinfo {author} {\bibfnamefont {Johan}\ \bibnamefont
  {Christensen}}, \ and\ \bibinfo {author} {\bibfnamefont {Martin}\
  \bibnamefont {van Hecke}},\ }\bibfield  {title} {\enquote {\bibinfo {title}
  {{Flexible mechanical metamaterials}},}\ }\href {\doibase
  10.1038/natrevmats.2017.66} {\bibfield  {journal} {\bibinfo  {journal}
  {Nature Reviews Materials}\ }\textbf {\bibinfo {volume} {2}},\ \bibinfo
  {pages} {17066} (\bibinfo {year} {2017})}\BibitemShut {NoStop}%
\bibitem [{\citenamefont {Overvelde}\ \emph {et~al.}(2017)\citenamefont
  {Overvelde}, \citenamefont {Weaver}, \citenamefont {Hoberman},\ and\
  \citenamefont {Bertoldi}}]{Overvelde2017}%
  \BibitemOpen
  \bibfield  {author} {\bibinfo {author} {\bibfnamefont {Johannes T.~B.}\
  \bibnamefont {Overvelde}}, \bibinfo {author} {\bibfnamefont {James~C.}\
  \bibnamefont {Weaver}}, \bibinfo {author} {\bibfnamefont {Chuck}\
  \bibnamefont {Hoberman}}, \ and\ \bibinfo {author} {\bibfnamefont {Katia}\
  \bibnamefont {Bertoldi}},\ }\bibfield  {title} {\enquote {\bibinfo {title}
  {{Rational design of reconfigurable prismatic architected materials}},}\
  }\href {\doibase 10.1038/nature20824} {\bibfield  {journal} {\bibinfo
  {journal} {Nature}\ }\textbf {\bibinfo {volume} {541}},\ \bibinfo {pages}
  {347--352} (\bibinfo {year} {2017})}\BibitemShut {NoStop}%
\bibitem [{\citenamefont {Gross}\ \emph {et~al.}(2019)\citenamefont {Gross},
  \citenamefont {Pantidis}, \citenamefont {Bertoldi},\ and\ \citenamefont
  {Gerasimidis}}]{Gross2019}%
  \BibitemOpen
  \bibfield  {author} {\bibinfo {author} {\bibfnamefont {Andrew}\ \bibnamefont
  {Gross}}, \bibinfo {author} {\bibfnamefont {Panos}\ \bibnamefont {Pantidis}},
  \bibinfo {author} {\bibfnamefont {Katia}\ \bibnamefont {Bertoldi}}, \ and\
  \bibinfo {author} {\bibfnamefont {Simos}\ \bibnamefont {Gerasimidis}},\
  }\bibfield  {title} {\enquote {\bibinfo {title} {{Correlation between
  topology and elastic properties of imperfect truss-lattice materials}},}\
  }\href {\doibase 10.1016/j.jmps.2018.11.007} {\bibfield  {journal} {\bibinfo
  {journal} {Journal of the Mechanics and Physics of Solids}\ }\textbf
  {\bibinfo {volume} {124}},\ \bibinfo {pages} {577--598} (\bibinfo {year}
  {2019})}\BibitemShut {NoStop}%
\bibitem [{\citenamefont {Goodrich}\ \emph {et~al.}(2015)\citenamefont
  {Goodrich}, \citenamefont {Liu},\ and\ \citenamefont {Nagel}}]{Goodrich2015}%
  \BibitemOpen
  \bibfield  {author} {\bibinfo {author} {\bibfnamefont {Carl~P.}\ \bibnamefont
  {Goodrich}}, \bibinfo {author} {\bibfnamefont {Andrea~J.}\ \bibnamefont
  {Liu}}, \ and\ \bibinfo {author} {\bibfnamefont {Sidney~R.}\ \bibnamefont
  {Nagel}},\ }\bibfield  {title} {\enquote {\bibinfo {title} {{The Principle of
  Independent Bond-Level Response: Tuning by Pruning to Exploit Disorder for
  Global Behavior}},}\ }\href {\doibase 10.1103/PhysRevLett.114.225501}
  {\bibfield  {journal} {\bibinfo  {journal} {Physical Review Letters}\
  }\textbf {\bibinfo {volume} {114}},\ \bibinfo {pages} {225501} (\bibinfo
  {year} {2015})}\BibitemShut {NoStop}%
\bibitem [{\citenamefont {Ronellenfitsch}\ \emph {et~al.}(2019)\citenamefont
  {Ronellenfitsch}, \citenamefont {Stoop}, \citenamefont {Yu}, \citenamefont
  {Forrow},\ and\ \citenamefont {Dunkel}}]{Ronellenfitsch2018}%
  \BibitemOpen
  \bibfield  {author} {\bibinfo {author} {\bibfnamefont {Henrik}\ \bibnamefont
  {Ronellenfitsch}}, \bibinfo {author} {\bibfnamefont {Norbert}\ \bibnamefont
  {Stoop}}, \bibinfo {author} {\bibfnamefont {Josephine}\ \bibnamefont {Yu}},
  \bibinfo {author} {\bibfnamefont {Aden}\ \bibnamefont {Forrow}}, \ and\
  \bibinfo {author} {\bibfnamefont {J{\"{o}}rn}\ \bibnamefont {Dunkel}},\
  }\bibfield  {title} {\enquote {\bibinfo {title} {{Inverse design of discrete
  mechanical metamaterials}},}\ }\href {\doibase
  10.1103/PhysRevMaterials.3.095201} {\bibfield  {journal} {\bibinfo  {journal}
  {Physical Review Materials}\ }\textbf {\bibinfo {volume} {3}},\ \bibinfo
  {pages} {095201} (\bibinfo {year} {2019})}\BibitemShut {NoStop}%
\bibitem [{\citenamefont {Gurtner}\ and\ \citenamefont
  {Durand}(2014)}]{Gurtner2014}%
  \BibitemOpen
  \bibfield  {author} {\bibinfo {author} {\bibfnamefont {G{\'{e}}rald}\
  \bibnamefont {Gurtner}}\ and\ \bibinfo {author} {\bibfnamefont {Marc}\
  \bibnamefont {Durand}},\ }\bibfield  {title} {\enquote {\bibinfo {title}
  {{Stiffest elastic networks}},}\ }\href {\doibase 10.1098/rspa.2013.0611}
  {\bibfield  {journal} {\bibinfo  {journal} {Proceedings of the Royal Society
  A: Mathematical, Physical and Engineering Sciences}\ }\textbf {\bibinfo
  {volume} {470}},\ \bibinfo {pages} {20130611} (\bibinfo {year}
  {2014})}\BibitemShut {NoStop}%
\bibitem [{\citenamefont {Katifori}(2018)}]{Katifori2018}%
  \BibitemOpen
  \bibfield  {author} {\bibinfo {author} {\bibfnamefont {Eleni}\ \bibnamefont
  {Katifori}},\ }\bibfield  {title} {\enquote {\bibinfo {title} {{The transport
  network of a leaf}},}\ }\href {\doibase 10.1016/j.crhy.2018.10.007}
  {\bibfield  {journal} {\bibinfo  {journal} {Comptes Rendus Physique}\
  }\textbf {\bibinfo {volume} {19}},\ \bibinfo {pages} {244--252} (\bibinfo
  {year} {2018})}\BibitemShut {NoStop}%
\bibitem [{\citenamefont {Ennos}(2005)}]{Ennos2005}%
  \BibitemOpen
  \bibfield  {author} {\bibinfo {author} {\bibfnamefont {A.R.}\ \bibnamefont
  {Ennos}},\ }\bibfield  {title} {\enquote {\bibinfo {title} {{Compliance in
  plants}},}\ }in\ \href {\doibase 10.2495/978-1-85312-941-4/02} {\emph
  {\bibinfo {booktitle} {Compliant Structures in Nature and Engineering}}},\
  Vol.~\bibinfo {volume} {20}\ (\bibinfo  {publisher} {WIT Press},\ \bibinfo
  {year} {2005})\ pp.\ \bibinfo {pages} {21--37}\BibitemShut {NoStop}%
\bibitem [{\citenamefont {Vogel}(2012)}]{Vogel2012}%
  \BibitemOpen
  \bibfield  {author} {\bibinfo {author} {\bibfnamefont {Steven}\ \bibnamefont
  {Vogel}},\ }\href {\doibase 10.7208/chicago/9780226859422.001.0001} {\emph
  {\bibinfo {title} {{The Life of a Leaf}}}}\ (\bibinfo  {publisher}
  {University of Chicago Press},\ \bibinfo {address} {Chicago, London},\
  \bibinfo {year} {2012})\BibitemShut {NoStop}%
\bibitem [{\citenamefont {Wright}\ \emph {et~al.}(2004)\citenamefont {Wright},
  \citenamefont {Reich}, \citenamefont {Westoby}, \citenamefont {Ackerly},
  \citenamefont {Baruch}, \citenamefont {Bongers}, \citenamefont
  {Cavender-Bares}, \citenamefont {Chapin}, \citenamefont {Cornelissen},
  \citenamefont {Diemer}, \citenamefont {Flexas}, \citenamefont {Garnier},
  \citenamefont {Groom}, \citenamefont {Gulias}, \citenamefont {Hikosaka},
  \citenamefont {Lamont}, \citenamefont {Lee}, \citenamefont {Lee},
  \citenamefont {Lusk}, \citenamefont {Midgley}, \citenamefont {Navas},
  \citenamefont {Niinemets}, \citenamefont {Oleksyn}, \citenamefont {Osada},
  \citenamefont {Poorter}, \citenamefont {Poot}, \citenamefont {Prior},
  \citenamefont {Pyankov}, \citenamefont {Roumet}, \citenamefont {Thomas},
  \citenamefont {Tjoelker}, \citenamefont {Veneklaas},\ and\ \citenamefont
  {Villar}}]{Wright2004}%
  \BibitemOpen
  \bibfield  {author} {\bibinfo {author} {\bibfnamefont {Ian~J}\ \bibnamefont
  {Wright}}, \bibinfo {author} {\bibfnamefont {Peter~B}\ \bibnamefont {Reich}},
  \bibinfo {author} {\bibfnamefont {Mark}\ \bibnamefont {Westoby}}, \bibinfo
  {author} {\bibfnamefont {David~D}\ \bibnamefont {Ackerly}}, \bibinfo {author}
  {\bibfnamefont {Zdravko}\ \bibnamefont {Baruch}}, \bibinfo {author}
  {\bibfnamefont {Frans}\ \bibnamefont {Bongers}}, \bibinfo {author}
  {\bibfnamefont {Jeannine}\ \bibnamefont {Cavender-Bares}}, \bibinfo {author}
  {\bibfnamefont {Terry}\ \bibnamefont {Chapin}}, \bibinfo {author}
  {\bibfnamefont {Johannes H~C}\ \bibnamefont {Cornelissen}}, \bibinfo {author}
  {\bibfnamefont {Matthias}\ \bibnamefont {Diemer}}, \bibinfo {author}
  {\bibfnamefont {Jaume}\ \bibnamefont {Flexas}}, \bibinfo {author}
  {\bibfnamefont {Eric}\ \bibnamefont {Garnier}}, \bibinfo {author}
  {\bibfnamefont {Philip~K}\ \bibnamefont {Groom}}, \bibinfo {author}
  {\bibfnamefont {Javier}\ \bibnamefont {Gulias}}, \bibinfo {author}
  {\bibfnamefont {Kouki}\ \bibnamefont {Hikosaka}}, \bibinfo {author}
  {\bibfnamefont {Byron~B}\ \bibnamefont {Lamont}}, \bibinfo {author}
  {\bibfnamefont {Tali}\ \bibnamefont {Lee}}, \bibinfo {author} {\bibfnamefont
  {William}\ \bibnamefont {Lee}}, \bibinfo {author} {\bibfnamefont
  {Christopher}\ \bibnamefont {Lusk}}, \bibinfo {author} {\bibfnamefont
  {Jeremy~J}\ \bibnamefont {Midgley}}, \bibinfo {author} {\bibfnamefont
  {Marie-Laure}\ \bibnamefont {Navas}}, \bibinfo {author} {\bibfnamefont
  {{\&}{\#}220;lo}\ \bibnamefont {Niinemets}}, \bibinfo {author} {\bibfnamefont
  {Jacek}\ \bibnamefont {Oleksyn}}, \bibinfo {author} {\bibfnamefont
  {Noriyuki}\ \bibnamefont {Osada}}, \bibinfo {author} {\bibfnamefont
  {Hendrik}\ \bibnamefont {Poorter}}, \bibinfo {author} {\bibfnamefont
  {Pieter}\ \bibnamefont {Poot}}, \bibinfo {author} {\bibfnamefont {Lynda}\
  \bibnamefont {Prior}}, \bibinfo {author} {\bibfnamefont {Vladimir~I}\
  \bibnamefont {Pyankov}}, \bibinfo {author} {\bibfnamefont {Catherine}\
  \bibnamefont {Roumet}}, \bibinfo {author} {\bibfnamefont {Sean~C}\
  \bibnamefont {Thomas}}, \bibinfo {author} {\bibfnamefont {Mark~G}\
  \bibnamefont {Tjoelker}}, \bibinfo {author} {\bibfnamefont {Erik~J}\
  \bibnamefont {Veneklaas}}, \ and\ \bibinfo {author} {\bibfnamefont {Rafael}\
  \bibnamefont {Villar}},\ }\bibfield  {title} {\enquote {\bibinfo {title}
  {{The worldwide leaf economics spectrum}},}\ }\href {\doibase
  10.1038/nature02403} {\bibfield  {journal} {\bibinfo  {journal} {Nature}\
  }\textbf {\bibinfo {volume} {428}},\ \bibinfo {pages} {821--827} (\bibinfo
  {year} {2004})}\BibitemShut {NoStop}%
\bibitem [{\citenamefont {Blonder}\ \emph {et~al.}(2011)\citenamefont
  {Blonder}, \citenamefont {Violle}, \citenamefont {Bentley},\ and\
  \citenamefont {Enquist}}]{Blonder2011}%
  \BibitemOpen
  \bibfield  {author} {\bibinfo {author} {\bibfnamefont {Benjamin}\
  \bibnamefont {Blonder}}, \bibinfo {author} {\bibfnamefont {Cyrille}\
  \bibnamefont {Violle}}, \bibinfo {author} {\bibfnamefont {Lisa~Patrick}\
  \bibnamefont {Bentley}}, \ and\ \bibinfo {author} {\bibfnamefont {Brian~J.}\
  \bibnamefont {Enquist}},\ }\bibfield  {title} {\enquote {\bibinfo {title}
  {{Venation networks and the origin of the leaf economics spectrum}},}\ }\href
  {\doibase 10.1111/j.1461-0248.2010.01554.x} {\bibfield  {journal} {\bibinfo
  {journal} {Ecology Letters}\ }\textbf {\bibinfo {volume} {14}},\ \bibinfo
  {pages} {91--100} (\bibinfo {year} {2011})}\BibitemShut {NoStop}%
\bibitem [{\citenamefont {Kull}\ and\ \citenamefont {Herbig}(1994)}]{Kull1994}%
  \BibitemOpen
  \bibfield  {author} {\bibinfo {author} {\bibfnamefont {Ulrich}\ \bibnamefont
  {Kull}}\ and\ \bibinfo {author} {\bibfnamefont {Astrid}\ \bibnamefont
  {Herbig}},\ }\bibfield  {title} {\enquote {\bibinfo {title} {{Leaf venation
  patterns and principles of evolution}},}\ }in\ \href {\doibase
  10.18419/opus-2304} {\emph {\bibinfo {booktitle} {Evolution of natural
  structures: Proceedings of the 3rd International Symposium of the
  Sonderforschungsbereich 230}}},\ \bibinfo {editor} {edited by\ \bibinfo
  {editor} {\bibfnamefont {Martin}\ \bibnamefont {Hilliges}}}\ (\bibinfo
  {publisher} {Vorstand des Sonderforschungsbereich 230, 1994 (Nat{\"{u}}rliche
  Konstruktionen 9)},\ \bibinfo {address} {Stuttgart},\ \bibinfo {year}
  {1994})\ pp.\ \bibinfo {pages} {167--175}\BibitemShut {NoStop}%
\bibitem [{\citenamefont {Katifori}\ and\ \citenamefont
  {Magnasco}(2012)}]{Katifori2012}%
  \BibitemOpen
  \bibfield  {author} {\bibinfo {author} {\bibfnamefont {Eleni}\ \bibnamefont
  {Katifori}}\ and\ \bibinfo {author} {\bibfnamefont {Marcelo~O.}\ \bibnamefont
  {Magnasco}},\ }\bibfield  {title} {\enquote {\bibinfo {title} {{Quantifying
  Loopy Network Architectures}},}\ }\href {\doibase
  10.1371/journal.pone.0037994} {\bibfield  {journal} {\bibinfo  {journal}
  {PLoS ONE}\ }\textbf {\bibinfo {volume} {7}},\ \bibinfo {pages} {e37994}
  (\bibinfo {year} {2012})}\BibitemShut {NoStop}%
\bibitem [{\citenamefont {Mileyko}\ \emph {et~al.}(2012)\citenamefont
  {Mileyko}, \citenamefont {Edelsbrunner}, \citenamefont {Price},\ and\
  \citenamefont {Weitz}}]{Mileyko2012}%
  \BibitemOpen
  \bibfield  {author} {\bibinfo {author} {\bibfnamefont {Yuriy}\ \bibnamefont
  {Mileyko}}, \bibinfo {author} {\bibfnamefont {Herbert}\ \bibnamefont
  {Edelsbrunner}}, \bibinfo {author} {\bibfnamefont {Charles~A.}\ \bibnamefont
  {Price}}, \ and\ \bibinfo {author} {\bibfnamefont {Joshua~S.}\ \bibnamefont
  {Weitz}},\ }\bibfield  {title} {\enquote {\bibinfo {title} {{Hierarchical
  Ordering of Reticular Networks}},}\ }\href {\doibase
  10.1371/journal.pone.0036715} {\bibfield  {journal} {\bibinfo  {journal}
  {PLoS ONE}\ }\textbf {\bibinfo {volume} {7}},\ \bibinfo {pages} {e36715}
  (\bibinfo {year} {2012})}\BibitemShut {NoStop}%
\bibitem [{\citenamefont {Ronellenfitsch}\ \emph {et~al.}(2015)\citenamefont
  {Ronellenfitsch}, \citenamefont {Lasser}, \citenamefont {Daly},\ and\
  \citenamefont {Katifori}}]{Ronellenfitsch2015b}%
  \BibitemOpen
  \bibfield  {author} {\bibinfo {author} {\bibfnamefont {Henrik}\ \bibnamefont
  {Ronellenfitsch}}, \bibinfo {author} {\bibfnamefont {Jana}\ \bibnamefont
  {Lasser}}, \bibinfo {author} {\bibfnamefont {Douglas~C.}\ \bibnamefont
  {Daly}}, \ and\ \bibinfo {author} {\bibfnamefont {Eleni}\ \bibnamefont
  {Katifori}},\ }\bibfield  {title} {\enquote {\bibinfo {title} {{Topological
  Phenotypes Constitute a New Dimension in the Phenotypic Space of Leaf
  Venation Networks}},}\ }\href {\doibase 10.1371/journal.pcbi.1004680}
  {\bibfield  {journal} {\bibinfo  {journal} {PLOS Computational Biology}\
  }\textbf {\bibinfo {volume} {11}},\ \bibinfo {pages} {e1004680} (\bibinfo
  {year} {2015})}\BibitemShut {NoStop}%
\bibitem [{\citenamefont {Katifori}\ \emph
  {et~al.}(2010{\natexlab{a}})\citenamefont {Katifori}, \citenamefont
  {Sz{\"{o}}llősi},\ and\ \citenamefont {Magnasco}}]{Katifori2010b}%
  \BibitemOpen
  \bibfield  {author} {\bibinfo {author} {\bibfnamefont {Eleni}\ \bibnamefont
  {Katifori}}, \bibinfo {author} {\bibfnamefont {Gergely~J.}\ \bibnamefont
  {Sz{\"{o}}llősi}}, \ and\ \bibinfo {author} {\bibfnamefont {Marcelo~O.}\
  \bibnamefont {Magnasco}},\ }\bibfield  {title} {\enquote {\bibinfo {title}
  {{Damage and Fluctuations Induce Loops in Optimal Transport Networks}},}\
  }\href {\doibase 10.1103/PhysRevLett.104.048704} {\bibfield  {journal}
  {\bibinfo  {journal} {Physical Review Letters}\ }\textbf {\bibinfo {volume}
  {104}},\ \bibinfo {pages} {048704} (\bibinfo {year}
  {2010}{\natexlab{a}})}\BibitemShut {NoStop}%
\bibitem [{\citenamefont {Corson}(2010)}]{Corson2010flow}%
  \BibitemOpen
  \bibfield  {author} {\bibinfo {author} {\bibfnamefont {Francis}\ \bibnamefont
  {Corson}},\ }\bibfield  {title} {\enquote {\bibinfo {title} {{Fluctuations
  and Redundancy in Optimal Transport Networks}},}\ }\href {\doibase
  10.1103/PhysRevLett.104.048703} {\bibfield  {journal} {\bibinfo  {journal}
  {Physical Review Letters}\ }\textbf {\bibinfo {volume} {104}},\ \bibinfo
  {pages} {048703} (\bibinfo {year} {2010})}\BibitemShut {NoStop}%
\bibitem [{\citenamefont {Hu}\ and\ \citenamefont {Cai}(2013)}]{Hu2013}%
  \BibitemOpen
  \bibfield  {author} {\bibinfo {author} {\bibfnamefont {Dan}\ \bibnamefont
  {Hu}}\ and\ \bibinfo {author} {\bibfnamefont {David}\ \bibnamefont {Cai}},\
  }\bibfield  {title} {\enquote {\bibinfo {title} {{Adaptation and Optimization
  of Biological Transport Networks}},}\ }\href {\doibase
  10.1103/PhysRevLett.111.138701} {\bibfield  {journal} {\bibinfo  {journal}
  {Physical Review Letters}\ }\textbf {\bibinfo {volume} {111}},\ \bibinfo
  {pages} {138701} (\bibinfo {year} {2013})}\BibitemShut {NoStop}%
\bibitem [{\citenamefont {Ronellenfitsch}\ and\ \citenamefont
  {Katifori}(2016)}]{Ronellenfitsch2016}%
  \BibitemOpen
  \bibfield  {author} {\bibinfo {author} {\bibfnamefont {Henrik}\ \bibnamefont
  {Ronellenfitsch}}\ and\ \bibinfo {author} {\bibfnamefont {Eleni}\
  \bibnamefont {Katifori}},\ }\bibfield  {title} {\enquote {\bibinfo {title}
  {{Global Optimization, Local Adaptation, and the Role of Growth in
  Distribution Networks}},}\ }\href {\doibase 10.1103/PhysRevLett.117.138301}
  {\bibfield  {journal} {\bibinfo  {journal} {Physical Review Letters}\
  }\textbf {\bibinfo {volume} {117}},\ \bibinfo {pages} {138301} (\bibinfo
  {year} {2016})}\BibitemShut {NoStop}%
\bibitem [{\citenamefont {Ronellenfitsch}\ and\ \citenamefont
  {Katifori}(2019)}]{Ronellenfitsch2019}%
  \BibitemOpen
  \bibfield  {author} {\bibinfo {author} {\bibfnamefont {Henrik}\ \bibnamefont
  {Ronellenfitsch}}\ and\ \bibinfo {author} {\bibfnamefont {Eleni}\
  \bibnamefont {Katifori}},\ }\bibfield  {title} {\enquote {\bibinfo {title}
  {{Phenotypes of Vascular Flow Networks}},}\ }\href {\doibase
  10.1103/PhysRevLett.123.248101} {\bibfield  {journal} {\bibinfo  {journal}
  {Physical Review Letters}\ }\textbf {\bibinfo {volume} {123}},\ \bibinfo
  {pages} {248101} (\bibinfo {year} {2019})}\BibitemShut {NoStop}%
\bibitem [{\citenamefont {Gavrilchenko}\ and\ \citenamefont
  {Katifori}(2019)}]{Gavrilchenko2019}%
  \BibitemOpen
  \bibfield  {author} {\bibinfo {author} {\bibfnamefont {Tatyana}\ \bibnamefont
  {Gavrilchenko}}\ and\ \bibinfo {author} {\bibfnamefont {Eleni}\ \bibnamefont
  {Katifori}},\ }\bibfield  {title} {\enquote {\bibinfo {title} {{Resilience in
  hierarchical fluid flow networks}},}\ }\href {\doibase
  10.1103/PhysRevE.99.012321} {\bibfield  {journal} {\bibinfo  {journal}
  {Physical Review E}\ }\textbf {\bibinfo {volume} {99}},\ \bibinfo {pages}
  {012321} (\bibinfo {year} {2019})}\BibitemShut {NoStop}%
\bibitem [{\citenamefont {Gilet}\ and\ \citenamefont
  {Bourouiba}(2015)}]{Gilet2015}%
  \BibitemOpen
  \bibfield  {author} {\bibinfo {author} {\bibfnamefont {T.}~\bibnamefont
  {Gilet}}\ and\ \bibinfo {author} {\bibfnamefont {L.}~\bibnamefont
  {Bourouiba}},\ }\bibfield  {title} {\enquote {\bibinfo {title} {{Fluid
  fragmentation shapes rain-induced foliar disease transmission}},}\ }\href
  {\doibase 10.1098/rsif.2014.1092} {\bibfield  {journal} {\bibinfo  {journal}
  {Journal of The Royal Society Interface}\ }\textbf {\bibinfo {volume} {12}},\
  \bibinfo {pages} {20141092} (\bibinfo {year} {2015})}\BibitemShut {NoStop}%
\bibitem [{\citenamefont {Wei}\ \emph {et~al.}(2012)\citenamefont {Wei},
  \citenamefont {Mandre},\ and\ \citenamefont {Mahadevan}}]{Wei2012}%
  \BibitemOpen
  \bibfield  {author} {\bibinfo {author} {\bibfnamefont {Z.}~\bibnamefont
  {Wei}}, \bibinfo {author} {\bibfnamefont {S.}~\bibnamefont {Mandre}}, \ and\
  \bibinfo {author} {\bibfnamefont {L.}~\bibnamefont {Mahadevan}},\ }\bibfield
  {title} {\enquote {\bibinfo {title} {{The branch with the furthest reach}},}\
  }\href {\doibase 10.1209/0295-5075/97/14005} {\bibfield  {journal} {\bibinfo
  {journal} {Europhysics Letters}\ }\textbf {\bibinfo {volume} {97}},\ \bibinfo
  {pages} {14005} (\bibinfo {year} {2012})}\BibitemShut {NoStop}%
\bibitem [{\citenamefont {Sun}\ \emph {et~al.}(2018)\citenamefont {Sun},
  \citenamefont {Cui}, \citenamefont {Zhu}, \citenamefont {Zhang},
  \citenamefont {Shi}, \citenamefont {Tang}, \citenamefont {Du}, \citenamefont
  {Liu}, \citenamefont {Cui}, \citenamefont {Chen},\ and\ \citenamefont
  {Guo}}]{Sun2018}%
  \BibitemOpen
  \bibfield  {author} {\bibinfo {author} {\bibfnamefont {Zhi}\ \bibnamefont
  {Sun}}, \bibinfo {author} {\bibfnamefont {Tianchen}\ \bibnamefont {Cui}},
  \bibinfo {author} {\bibfnamefont {Yichao}\ \bibnamefont {Zhu}}, \bibinfo
  {author} {\bibfnamefont {Weisheng}\ \bibnamefont {Zhang}}, \bibinfo {author}
  {\bibfnamefont {Shanshan}\ \bibnamefont {Shi}}, \bibinfo {author}
  {\bibfnamefont {Shan}\ \bibnamefont {Tang}}, \bibinfo {author} {\bibfnamefont
  {Zongliang}\ \bibnamefont {Du}}, \bibinfo {author} {\bibfnamefont {Chang}\
  \bibnamefont {Liu}}, \bibinfo {author} {\bibfnamefont {Ronghua}\ \bibnamefont
  {Cui}}, \bibinfo {author} {\bibfnamefont {Hongjie}\ \bibnamefont {Chen}}, \
  and\ \bibinfo {author} {\bibfnamefont {Xu}~\bibnamefont {Guo}},\ }\bibfield
  {title} {\enquote {\bibinfo {title} {{The mechanical principles behind the
  golden ratio distribution of veins in plant leaves}},}\ }\href {\doibase
  10.1038/s41598-018-31763-1} {\bibfield  {journal} {\bibinfo  {journal}
  {Scientific Reports}\ }\textbf {\bibinfo {volume} {8}},\ \bibinfo {pages}
  {13859} (\bibinfo {year} {2018})}\BibitemShut {NoStop}%
\bibitem [{\citenamefont {Blonder}\ \emph {et~al.}(2020)\citenamefont
  {Blonder}, \citenamefont {Both}, \citenamefont {Jodra}, \citenamefont {Xu},
  \citenamefont {Fricker}, \citenamefont {Matos}, \citenamefont {Majalap},
  \citenamefont {Burslem}, \citenamefont {Teh},\ and\ \citenamefont
  {Malhi}}]{Blonder2020}%
  \BibitemOpen
  \bibfield  {author} {\bibinfo {author} {\bibfnamefont {Benjamin}\
  \bibnamefont {Blonder}}, \bibinfo {author} {\bibfnamefont {Sabine}\
  \bibnamefont {Both}}, \bibinfo {author} {\bibfnamefont {Miguel}\ \bibnamefont
  {Jodra}}, \bibinfo {author} {\bibfnamefont {Hao}\ \bibnamefont {Xu}},
  \bibinfo {author} {\bibfnamefont {Mark}\ \bibnamefont {Fricker}}, \bibinfo
  {author} {\bibfnamefont {Ila{\'{i}}ne~S.}\ \bibnamefont {Matos}}, \bibinfo
  {author} {\bibfnamefont {Noreen}\ \bibnamefont {Majalap}}, \bibinfo {author}
  {\bibfnamefont {David~F.R.P.}\ \bibnamefont {Burslem}}, \bibinfo {author}
  {\bibfnamefont {Yit~Arn}\ \bibnamefont {Teh}}, \ and\ \bibinfo {author}
  {\bibfnamefont {Yadvinder}\ \bibnamefont {Malhi}},\ }\bibfield  {title}
  {\enquote {\bibinfo {title} {{Linking functional traits to multiscale
  statistics of leaf venation networks}},}\ }\href {\doibase 10.1111/nph.16830}
  {\bibfield  {journal} {\bibinfo  {journal} {New Phytologist}\ }\textbf
  {\bibinfo {volume} {228}},\ \bibinfo {pages} {1796--1810} (\bibinfo {year}
  {2020})}\BibitemShut {NoStop}%
\bibitem [{\citenamefont {Bohn}\ and\ \citenamefont
  {Magnasco}(2007)}]{Bohn2007}%
  \BibitemOpen
  \bibfield  {author} {\bibinfo {author} {\bibfnamefont {Steffen}\ \bibnamefont
  {Bohn}}\ and\ \bibinfo {author} {\bibfnamefont {Marcelo~O.}\ \bibnamefont
  {Magnasco}},\ }\bibfield  {title} {\enquote {\bibinfo {title} {{Structure,
  Scaling, and Phase Transition in the Optimal Transport Network}},}\ }\href
  {\doibase 10.1103/PhysRevLett.98.088702} {\bibfield  {journal} {\bibinfo
  {journal} {Physical Review Letters}\ }\textbf {\bibinfo {volume} {98}},\
  \bibinfo {pages} {088702} (\bibinfo {year} {2007})}\BibitemShut {NoStop}%
\bibitem [{\citenamefont {Durand}(2007)}]{Durand2007}%
  \BibitemOpen
  \bibfield  {author} {\bibinfo {author} {\bibfnamefont {Marc}\ \bibnamefont
  {Durand}},\ }\bibfield  {title} {\enquote {\bibinfo {title} {{Structure of
  Optimal Transport Networks Subject to a Global Constraint}},}\ }\href
  {\doibase 10.1103/PhysRevLett.98.088701} {\bibfield  {journal} {\bibinfo
  {journal} {Physical Review Letters}\ }\textbf {\bibinfo {volume} {98}},\
  \bibinfo {pages} {088701} (\bibinfo {year} {2007})}\BibitemShut {NoStop}%
\bibitem [{\citenamefont {Savage}\ \emph {et~al.}(2010)\citenamefont {Savage},
  \citenamefont {Bentley}, \citenamefont {Enquist}, \citenamefont {Sperry},
  \citenamefont {Smith}, \citenamefont {Reich},\ and\ \citenamefont {von
  Allmen}}]{Savage2010}%
  \BibitemOpen
  \bibfield  {author} {\bibinfo {author} {\bibfnamefont {V~M}\ \bibnamefont
  {Savage}}, \bibinfo {author} {\bibfnamefont {L~P}\ \bibnamefont {Bentley}},
  \bibinfo {author} {\bibfnamefont {B~J}\ \bibnamefont {Enquist}}, \bibinfo
  {author} {\bibfnamefont {J~S}\ \bibnamefont {Sperry}}, \bibinfo {author}
  {\bibfnamefont {D~D}\ \bibnamefont {Smith}}, \bibinfo {author} {\bibfnamefont
  {P~B}\ \bibnamefont {Reich}}, \ and\ \bibinfo {author} {\bibfnamefont {E~I}\
  \bibnamefont {von Allmen}},\ }\bibfield  {title} {\enquote {\bibinfo {title}
  {{Hydraulic trade-offs and space filling enable better predictions of
  vascular structure and function in plants}},}\ }\href {\doibase
  10.1073/pnas.1012194108} {\bibfield  {journal} {\bibinfo  {journal}
  {Proceedings of the National Academy of Sciences}\ }\textbf {\bibinfo
  {volume} {107}},\ \bibinfo {pages} {22722--22727} (\bibinfo {year}
  {2010})}\BibitemShut {NoStop}%
\bibitem [{\citenamefont {Price}\ and\ \citenamefont
  {Weitz}(2014)}]{Price2014b}%
  \BibitemOpen
  \bibfield  {author} {\bibinfo {author} {\bibfnamefont {Charles~A.}\
  \bibnamefont {Price}}\ and\ \bibinfo {author} {\bibfnamefont {Joshua~S.}\
  \bibnamefont {Weitz}},\ }\bibfield  {title} {\enquote {\bibinfo {title}
  {{Costs and benefits of reticulate leaf venation}},}\ }\href {\doibase
  10.1186/s12870-014-0234-2} {\bibfield  {journal} {\bibinfo  {journal} {BMC
  Plant Biology}\ }\textbf {\bibinfo {volume} {14}},\ \bibinfo {pages} {234}
  (\bibinfo {year} {2014})}\BibitemShut {NoStop}%
\bibitem [{\citenamefont {Murray}(1926)}]{Murray1926}%
  \BibitemOpen
  \bibfield  {author} {\bibinfo {author} {\bibfnamefont {Cecil~D.}\
  \bibnamefont {Murray}},\ }\bibfield  {title} {\enquote {\bibinfo {title}
  {{The Physiological Principle of Minimum Work: I. The Vascular System and the
  Cost of Blood Volume}},}\ }\href {\doibase 10.1073/pnas.12.3.207} {\bibfield
  {journal} {\bibinfo  {journal} {Proceedings of the National Academy of
  Sciences}\ }\textbf {\bibinfo {volume} {12}},\ \bibinfo {pages} {207--214}
  (\bibinfo {year} {1926})}\BibitemShut {NoStop}%
\bibitem [{\citenamefont {West}\ \emph {et~al.}(1999)\citenamefont {West},
  \citenamefont {Brown},\ and\ \citenamefont {Enquist}}]{West1999}%
  \BibitemOpen
  \bibfield  {author} {\bibinfo {author} {\bibfnamefont {Geoffrey~B.}\
  \bibnamefont {West}}, \bibinfo {author} {\bibfnamefont {James~H.}\
  \bibnamefont {Brown}}, \ and\ \bibinfo {author} {\bibfnamefont {Brian~J.}\
  \bibnamefont {Enquist}},\ }\bibfield  {title} {\enquote {\bibinfo {title} {{A
  general model for the structure and allometry of plant vascular systems}},}\
  }\href {\doibase 10.1038/23251} {\bibfield  {journal} {\bibinfo  {journal}
  {Nature}\ }\textbf {\bibinfo {volume} {400}},\ \bibinfo {pages} {664--667}
  (\bibinfo {year} {1999})}\BibitemShut {NoStop}%
\bibitem [{\citenamefont {Kirkegaard}\ and\ \citenamefont
  {Sneppen}(2020)}]{Kirkegaard2020}%
  \BibitemOpen
  \bibfield  {author} {\bibinfo {author} {\bibfnamefont {Julius~B.}\
  \bibnamefont {Kirkegaard}}\ and\ \bibinfo {author} {\bibfnamefont {Kim}\
  \bibnamefont {Sneppen}},\ }\bibfield  {title} {\enquote {\bibinfo {title}
  {Optimal transport flows for distributed production networks},}\ }\href
  {\doibase 10.1103/PhysRevLett.124.208101} {\bibfield  {journal} {\bibinfo
  {journal} {Physical Review Letters}\ }\textbf {\bibinfo {volume} {124}},\
  \bibinfo {pages} {208101} (\bibinfo {year} {2020})}\BibitemShut {NoStop}%
\bibitem [{\citenamefont {Safran}(1999)}]{Safran1999}%
  \BibitemOpen
  \bibfield  {author} {\bibinfo {author} {\bibfnamefont {S.~A.}\ \bibnamefont
  {Safran}},\ }\bibfield  {title} {\enquote {\bibinfo {title} {{Curvature
  elasticity of thin films}},}\ }\href {\doibase 10.1080/000187399243428}
  {\bibfield  {journal} {\bibinfo  {journal} {Advances in Physics}\ }\textbf
  {\bibinfo {volume} {48}},\ \bibinfo {pages} {395--448} (\bibinfo {year}
  {1999})}\BibitemShut {NoStop}%
\bibitem [{\citenamefont {Helfrich}(1973)}]{Helfrich1973}%
  \BibitemOpen
  \bibfield  {author} {\bibinfo {author} {\bibfnamefont {W.}~\bibnamefont
  {Helfrich}},\ }\bibfield  {title} {\enquote {\bibinfo {title} {{Elastic
  Properties of Lipid Bilayers: Theory and Possible Experiments}},}\ }\href
  {\doibase 10.1515/znc-1973-11-1209} {\bibfield  {journal} {\bibinfo
  {journal} {Zeitschrift f{\"{u}}r Naturforschung C}\ }\textbf {\bibinfo
  {volume} {28}},\ \bibinfo {pages} {693--703} (\bibinfo {year}
  {1973})}\BibitemShut {NoStop}%
\bibitem [{\citenamefont {Katifori}\ \emph
  {et~al.}(2010{\natexlab{b}})\citenamefont {Katifori}, \citenamefont {Alben},
  \citenamefont {Cerda}, \citenamefont {Nelson},\ and\ \citenamefont
  {Dumais}}]{Katifori2010}%
  \BibitemOpen
  \bibfield  {author} {\bibinfo {author} {\bibfnamefont {Eleni}\ \bibnamefont
  {Katifori}}, \bibinfo {author} {\bibfnamefont {Silas}\ \bibnamefont {Alben}},
  \bibinfo {author} {\bibfnamefont {Enrique}\ \bibnamefont {Cerda}}, \bibinfo
  {author} {\bibfnamefont {David~R.}\ \bibnamefont {Nelson}}, \ and\ \bibinfo
  {author} {\bibfnamefont {Jacques}\ \bibnamefont {Dumais}},\ }\bibfield
  {title} {\enquote {\bibinfo {title} {{Foldable structures and the natural
  design of pollen grains}},}\ }\href {\doibase 10.1073/pnas.0911223107}
  {\bibfield  {journal} {\bibinfo  {journal} {Proceedings of the National
  Academy of Sciences}\ }\textbf {\bibinfo {volume} {107}},\ \bibinfo {pages}
  {7635--7639} (\bibinfo {year} {2010}{\natexlab{b}})}\BibitemShut {NoStop}%
\bibitem [{\citenamefont {Couturier}\ \emph {et~al.}(2013)\citenamefont
  {Couturier}, \citenamefont {Dumais}, \citenamefont {Cerda},\ and\
  \citenamefont {Katifori}}]{Couturier2013}%
  \BibitemOpen
  \bibfield  {author} {\bibinfo {author} {\bibfnamefont {E.}~\bibnamefont
  {Couturier}}, \bibinfo {author} {\bibfnamefont {J.}~\bibnamefont {Dumais}},
  \bibinfo {author} {\bibfnamefont {E.}~\bibnamefont {Cerda}}, \ and\ \bibinfo
  {author} {\bibfnamefont {E.}~\bibnamefont {Katifori}},\ }\bibfield  {title}
  {\enquote {\bibinfo {title} {{Folding of an opened spherical shell}},}\
  }\href {\doibase 10.1039/c3sm50575h} {\bibfield  {journal} {\bibinfo
  {journal} {Soft Matter}\ }\textbf {\bibinfo {volume} {9}},\ \bibinfo {pages}
  {8359--8367} (\bibinfo {year} {2013})}\BibitemShut {NoStop}%
\bibitem [{\citenamefont {Seung}\ and\ \citenamefont
  {Nelson}(1988)}]{Seung1988}%
  \BibitemOpen
  \bibfield  {author} {\bibinfo {author} {\bibfnamefont {H.~S.}\ \bibnamefont
  {Seung}}\ and\ \bibinfo {author} {\bibfnamefont {David~R}\ \bibnamefont
  {Nelson}},\ }\bibfield  {title} {\enquote {\bibinfo {title} {{Defects in
  flexible membranes with crystalline order}},}\ }\href {\doibase
  10.1103/PhysRevA.38.1005} {\bibfield  {journal} {\bibinfo  {journal}
  {Physical Review A}\ }\textbf {\bibinfo {volume} {38}},\ \bibinfo {pages}
  {1005--1018} (\bibinfo {year} {1988})}\BibitemShut {NoStop}%
\bibitem [{\citenamefont {Liang}\ and\ \citenamefont
  {Mahadevan}(2009)}]{Liang2009}%
  \BibitemOpen
  \bibfield  {author} {\bibinfo {author} {\bibfnamefont {Haiyi}\ \bibnamefont
  {Liang}}\ and\ \bibinfo {author} {\bibfnamefont {L.}~\bibnamefont
  {Mahadevan}},\ }\bibfield  {title} {\enquote {\bibinfo {title} {{The shape of
  a long leaf}},}\ }\href {\doibase 10.1073/pnas.0911954106} {\bibfield
  {journal} {\bibinfo  {journal} {Proceedings of the National Academy of
  Sciences}\ }\textbf {\bibinfo {volume} {106}},\ \bibinfo {pages}
  {22049--22054} (\bibinfo {year} {2009})}\BibitemShut {NoStop}%
\bibitem [{\citenamefont {Guckenberger}\ \emph {et~al.}(2016)\citenamefont
  {Guckenberger}, \citenamefont {Schraml}, \citenamefont {Chen}, \citenamefont
  {Leonetti},\ and\ \citenamefont {Gekle}}]{Guckenberger2016}%
  \BibitemOpen
  \bibfield  {author} {\bibinfo {author} {\bibfnamefont {Achim}\ \bibnamefont
  {Guckenberger}}, \bibinfo {author} {\bibfnamefont {Marcel~P.}\ \bibnamefont
  {Schraml}}, \bibinfo {author} {\bibfnamefont {Paul~G.}\ \bibnamefont {Chen}},
  \bibinfo {author} {\bibfnamefont {Marc}\ \bibnamefont {Leonetti}}, \ and\
  \bibinfo {author} {\bibfnamefont {Stephan}\ \bibnamefont {Gekle}},\
  }\bibfield  {title} {\enquote {\bibinfo {title} {{On the bending algorithms
  for soft objects in flows}},}\ }\href {\doibase 10.1016/j.cpc.2016.04.018}
  {\bibfield  {journal} {\bibinfo  {journal} {Computer Physics Communications}\
  }\textbf {\bibinfo {volume} {207}},\ \bibinfo {pages} {1--23} (\bibinfo
  {year} {2016})}\BibitemShut {NoStop}%
\bibitem [{\citenamefont {Gompper}\ and\ \citenamefont
  {Kroll}(1996)}]{Gompper1996}%
  \BibitemOpen
  \bibfield  {author} {\bibinfo {author} {\bibfnamefont {G.}~\bibnamefont
  {Gompper}}\ and\ \bibinfo {author} {\bibfnamefont {D.~M.}\ \bibnamefont
  {Kroll}},\ }\bibfield  {title} {\enquote {\bibinfo {title} {{Random Surface
  Discretizations and the Renormalization of the Bending Rigidity}},}\ }\href
  {\doibase 10.1051/jp1:1996246} {\bibfield  {journal} {\bibinfo  {journal}
  {Journal de Physique I}\ }\textbf {\bibinfo {volume} {6}},\ \bibinfo {pages}
  {1305--1320} (\bibinfo {year} {1996})}\BibitemShut {NoStop}%
\bibitem [{\citenamefont {Witten}(2007)}]{Witten2007}%
  \BibitemOpen
  \bibfield  {author} {\bibinfo {author} {\bibfnamefont {T.~A.}\ \bibnamefont
  {Witten}},\ }\bibfield  {title} {\enquote {\bibinfo {title} {{Stress focusing
  in elastic sheets}},}\ }\href {\doibase 10.1103/RevModPhys.79.643} {\bibfield
   {journal} {\bibinfo  {journal} {Reviews of Modern Physics}\ }\textbf
  {\bibinfo {volume} {79}},\ \bibinfo {pages} {643--675} (\bibinfo {year}
  {2007})}\BibitemShut {NoStop}%
\bibitem [{\citenamefont {Audoly}\ and\ \citenamefont
  {Pomeau}(2010)}]{Audoly2010}%
  \BibitemOpen
  \bibfield  {author} {\bibinfo {author} {\bibfnamefont {Basile}\ \bibnamefont
  {Audoly}}\ and\ \bibinfo {author} {\bibfnamefont {Yves}\ \bibnamefont
  {Pomeau}},\ }\href@noop {} {\emph {\bibinfo {title} {{Elasticity and
  Geometry}}}}\ (\bibinfo  {publisher} {Oxford University Press},\ \bibinfo
  {address} {Oxford},\ \bibinfo {year} {2010})\ p.\ \bibinfo {pages}
  {586}\BibitemShut {NoStop}%
\bibitem [{\citenamefont {Bergou}\ \emph {et~al.}(2008)\citenamefont {Bergou},
  \citenamefont {Wardetzky}, \citenamefont {Robinson}, \citenamefont {Audoly},\
  and\ \citenamefont {Grinspun}}]{Bergou2008}%
  \BibitemOpen
  \bibfield  {author} {\bibinfo {author} {\bibfnamefont {Mikl{\'{o}}s}\
  \bibnamefont {Bergou}}, \bibinfo {author} {\bibfnamefont {Max}\ \bibnamefont
  {Wardetzky}}, \bibinfo {author} {\bibfnamefont {Stephen}\ \bibnamefont
  {Robinson}}, \bibinfo {author} {\bibfnamefont {Basile}\ \bibnamefont
  {Audoly}}, \ and\ \bibinfo {author} {\bibfnamefont {Eitan}\ \bibnamefont
  {Grinspun}},\ }\bibfield  {title} {\enquote {\bibinfo {title} {{Discrete
  elastic rods}},}\ }\href {\doibase 10.1145/1360612.1360662} {\bibfield
  {journal} {\bibinfo  {journal} {ACM Transactions on Graphics}\ }\textbf
  {\bibinfo {volume} {27}},\ \bibinfo {pages} {1--12} (\bibinfo {year}
  {2008})}\BibitemShut {NoStop}%
\bibitem [{Note1()}]{Note1}%
  \BibitemOpen
  \bibinfo {note} {See Supplemental Material [url] for a detailed discussion of
  the approximations, a derivation the elastic energy and the constrained
  optimization algorithm, the continuum limit, a discussion of mechanical
  constraints and optimization under self-loads, three-dimensional DBNs, and a
  comparison of optimal DBNs to real leaf networks using topological metrics,
  which includes Refs.~\cite
  {DoCarmo1988,Lubensky2015,Bruyneel2005}}\BibitemShut {NoStop}%
\bibitem [{\citenamefont {Chartrand}(2007)}]{Chartrand2007}%
  \BibitemOpen
  \bibfield  {author} {\bibinfo {author} {\bibfnamefont {Rick}\ \bibnamefont
  {Chartrand}},\ }\bibfield  {title} {\enquote {\bibinfo {title} {{Exact
  Reconstruction of Sparse Signals via Nonconvex Minimization}},}\ }\href
  {\doibase 10.1109/LSP.2007.898300} {\bibfield  {journal} {\bibinfo  {journal}
  {IEEE Signal Processing Letters}\ }\textbf {\bibinfo {volume} {14}},\
  \bibinfo {pages} {707--710} (\bibinfo {year} {2007})}\BibitemShut {NoStop}%
\bibitem [{\citenamefont {John}\ \emph {et~al.}(2017)\citenamefont {John},
  \citenamefont {Scoffoni}, \citenamefont {Buckley}, \citenamefont {Villar},
  \citenamefont {Poorter},\ and\ \citenamefont {Sack}}]{John2017}%
  \BibitemOpen
  \bibfield  {author} {\bibinfo {author} {\bibfnamefont {Grace~P.}\
  \bibnamefont {John}}, \bibinfo {author} {\bibfnamefont {Christine}\
  \bibnamefont {Scoffoni}}, \bibinfo {author} {\bibfnamefont {Thomas~N.}\
  \bibnamefont {Buckley}}, \bibinfo {author} {\bibfnamefont {Rafael}\
  \bibnamefont {Villar}}, \bibinfo {author} {\bibfnamefont {Hendrik}\
  \bibnamefont {Poorter}}, \ and\ \bibinfo {author} {\bibfnamefont {Lawren}\
  \bibnamefont {Sack}},\ }\bibfield  {title} {\enquote {\bibinfo {title} {{The
  anatomical and compositional basis of leaf mass per area}},}\ }\href
  {\doibase 10.1111/ele.12739} {\bibfield  {journal} {\bibinfo  {journal}
  {Ecology Letters}\ }\textbf {\bibinfo {volume} {20}},\ \bibinfo {pages}
  {412--425} (\bibinfo {year} {2017})}\BibitemShut {NoStop}%
\bibitem [{\citenamefont {Banavar}\ \emph {et~al.}(2000)\citenamefont
  {Banavar}, \citenamefont {Colaiori}, \citenamefont {Flammini}, \citenamefont
  {Maritan},\ and\ \citenamefont {Rinaldo}}]{Banavar2000}%
  \BibitemOpen
  \bibfield  {author} {\bibinfo {author} {\bibfnamefont {Jayanth~R.}\
  \bibnamefont {Banavar}}, \bibinfo {author} {\bibfnamefont {Francesca}\
  \bibnamefont {Colaiori}}, \bibinfo {author} {\bibfnamefont {Alessandro}\
  \bibnamefont {Flammini}}, \bibinfo {author} {\bibfnamefont {Amos}\
  \bibnamefont {Maritan}}, \ and\ \bibinfo {author} {\bibfnamefont {Andrea}\
  \bibnamefont {Rinaldo}},\ }\bibfield  {title} {\enquote {\bibinfo {title}
  {{Topology of the Fittest Transportation Network}},}\ }\href {\doibase
  10.1103/PhysRevLett.84.4745} {\bibfield  {journal} {\bibinfo  {journal}
  {Physical Review Letters}\ }\textbf {\bibinfo {volume} {84}},\ \bibinfo
  {pages} {4745--4748} (\bibinfo {year} {2000})}\BibitemShut {NoStop}%
\bibitem [{\citenamefont {Yan}\ \emph {et~al.}(2018)\citenamefont {Yan},
  \citenamefont {Wang},\ and\ \citenamefont {Sigmund}}]{Yan2018}%
  \BibitemOpen
  \bibfield  {author} {\bibinfo {author} {\bibfnamefont {Suna}\ \bibnamefont
  {Yan}}, \bibinfo {author} {\bibfnamefont {Fengwen}\ \bibnamefont {Wang}}, \
  and\ \bibinfo {author} {\bibfnamefont {Ole}\ \bibnamefont {Sigmund}},\
  }\bibfield  {title} {\enquote {\bibinfo {title} {{On the non-optimality of
  tree structures for heat conduction}},}\ }\href {\doibase
  10.1016/j.ijheatmasstransfer.2018.01.114} {\bibfield  {journal} {\bibinfo
  {journal} {International Journal of Heat and Mass Transfer}\ }\textbf
  {\bibinfo {volume} {122}},\ \bibinfo {pages} {660--680} (\bibinfo {year}
  {2018})}\BibitemShut {NoStop}%
\bibitem [{\citenamefont {Runions}\ \emph {et~al.}(2014)\citenamefont
  {Runions}, \citenamefont {Smith},\ and\ \citenamefont
  {Prusinkiewicz}}]{Runions2014}%
  \BibitemOpen
  \bibfield  {author} {\bibinfo {author} {\bibfnamefont {Adam}\ \bibnamefont
  {Runions}}, \bibinfo {author} {\bibfnamefont {Richard~S.}\ \bibnamefont
  {Smith}}, \ and\ \bibinfo {author} {\bibfnamefont {Przemyslaw}\ \bibnamefont
  {Prusinkiewicz}},\ }\bibfield  {title} {\enquote {\bibinfo {title}
  {Computational models of auxin-driven development},}\ }in\ \href {\doibase
  10.1007/978-3-7091-1526-8_15} {\emph {\bibinfo {booktitle} {{Auxin and Its
  Role in Plant Development}}}}\ (\bibinfo  {publisher} {Springer-Verlag
  Wien},\ \bibinfo {year} {2014})\ pp.\ \bibinfo {pages} {315--357}\BibitemShut
  {NoStop}%
\bibitem [{\citenamefont {Dimitrov}\ and\ \citenamefont
  {Zucker}(2006)}]{Dimitrov2006}%
  \BibitemOpen
  \bibfield  {author} {\bibinfo {author} {\bibfnamefont {Pavel}\ \bibnamefont
  {Dimitrov}}\ and\ \bibinfo {author} {\bibfnamefont {Steven~W}\ \bibnamefont
  {Zucker}},\ }\bibfield  {title} {\enquote {\bibinfo {title} {{A constant
  production hypothesis guides leaf venation patterning}},}\ }\href {\doibase
  10.1073/pnas.0603559103} {\bibfield  {journal} {\bibinfo  {journal}
  {Proceedings of the National Academy of Sciences}\ }\textbf {\bibinfo
  {volume} {103}},\ \bibinfo {pages} {9363--9368} (\bibinfo {year}
  {2006})}\BibitemShut {NoStop}%
\bibitem [{\citenamefont {Corson}\ \emph {et~al.}(2010)\citenamefont {Corson},
  \citenamefont {Henry},\ and\ \citenamefont
  {Adda-Bedia}}]{Corson2010mechanics}%
  \BibitemOpen
  \bibfield  {author} {\bibinfo {author} {\bibfnamefont {F.}~\bibnamefont
  {Corson}}, \bibinfo {author} {\bibfnamefont {H.}~\bibnamefont {Henry}}, \
  and\ \bibinfo {author} {\bibfnamefont {M.}~\bibnamefont {Adda-Bedia}},\
  }\bibfield  {title} {\enquote {\bibinfo {title} {{A model for hierarchical
  patterns under mechanical stresses}},}\ }\href {\doibase
  10.1080/14786430903196665} {\bibfield  {journal} {\bibinfo  {journal}
  {Philosophical Magazine}\ }\textbf {\bibinfo {volume} {90}},\ \bibinfo
  {pages} {357--373} (\bibinfo {year} {2010})}\BibitemShut {NoStop}%
\bibitem [{\citenamefont {Corson}\ \emph {et~al.}(2009)\citenamefont {Corson},
  \citenamefont {Adda-Bedia},\ and\ \citenamefont {Boudaoud}}]{Corson2009}%
  \BibitemOpen
  \bibfield  {author} {\bibinfo {author} {\bibfnamefont {Francis}\ \bibnamefont
  {Corson}}, \bibinfo {author} {\bibfnamefont {Mokhtar}\ \bibnamefont
  {Adda-Bedia}}, \ and\ \bibinfo {author} {\bibfnamefont {Arezki}\ \bibnamefont
  {Boudaoud}},\ }\bibfield  {title} {\enquote {\bibinfo {title} {{In silico
  leaf venation networks: Growth and reorganization driven by mechanical
  forces}},}\ }\href {\doibase 10.1016/j.jtbi.2009.05.002} {\bibfield
  {journal} {\bibinfo  {journal} {Journal of Theoretical Biology}\ }\textbf
  {\bibinfo {volume} {259}},\ \bibinfo {pages} {440--448} (\bibinfo {year}
  {2009})}\BibitemShut {NoStop}%
\bibitem [{\citenamefont {Laguna}\ \emph {et~al.}(2008)\citenamefont {Laguna},
  \citenamefont {Bohn},\ and\ \citenamefont {Jagla}}]{Laguna2008}%
  \BibitemOpen
  \bibfield  {author} {\bibinfo {author} {\bibfnamefont {Maria~F.}\
  \bibnamefont {Laguna}}, \bibinfo {author} {\bibfnamefont {Steffen}\
  \bibnamefont {Bohn}}, \ and\ \bibinfo {author} {\bibfnamefont {Eduardo~A.}\
  \bibnamefont {Jagla}},\ }\bibfield  {title} {\enquote {\bibinfo {title} {{The
  Role of Elastic Stresses on Leaf Venation Morphogenesis}},}\ }\href {\doibase
  10.1371/journal.pcbi.1000055} {\bibfield  {journal} {\bibinfo  {journal}
  {PLoS Computational Biology}\ }\textbf {\bibinfo {volume} {4}},\ \bibinfo
  {pages} {e1000055} (\bibinfo {year} {2008})}\BibitemShut {NoStop}%
\bibitem [{\citenamefont {Bar-Sinai}\ \emph {et~al.}(2016)\citenamefont
  {Bar-Sinai}, \citenamefont {Julien}, \citenamefont {Sharon}, \citenamefont
  {Armon}, \citenamefont {Nakayama}, \citenamefont {Adda-Bedia},\ and\
  \citenamefont {Boudaoud}}]{Bar-Sinai2016}%
  \BibitemOpen
  \bibfield  {author} {\bibinfo {author} {\bibfnamefont {Yohai}\ \bibnamefont
  {Bar-Sinai}}, \bibinfo {author} {\bibfnamefont {Jean-Daniel}\ \bibnamefont
  {Julien}}, \bibinfo {author} {\bibfnamefont {Eran}\ \bibnamefont {Sharon}},
  \bibinfo {author} {\bibfnamefont {Shahaf}\ \bibnamefont {Armon}}, \bibinfo
  {author} {\bibfnamefont {Naomi}\ \bibnamefont {Nakayama}}, \bibinfo {author}
  {\bibfnamefont {Mokhtar}\ \bibnamefont {Adda-Bedia}}, \ and\ \bibinfo
  {author} {\bibfnamefont {Arezki}\ \bibnamefont {Boudaoud}},\ }\bibfield
  {title} {\enquote {\bibinfo {title} {{Mechanical Stress Induces Remodeling of
  Vascular Networks in Growing Leaves}},}\ }\href {\doibase
  10.1371/journal.pcbi.1004819} {\bibfield  {journal} {\bibinfo  {journal}
  {PLOS Computational Biology}\ }\textbf {\bibinfo {volume} {12}},\ \bibinfo
  {pages} {e1004819} (\bibinfo {year} {2016})}\BibitemShut {NoStop}%
\bibitem [{\citenamefont {Couder}\ \emph {et~al.}(2002)\citenamefont {Couder},
  \citenamefont {Pauchard}, \citenamefont {Allain}, \citenamefont
  {Adda-Bedia},\ and\ \citenamefont {Douady}}]{Couder2002}%
  \BibitemOpen
  \bibfield  {author} {\bibinfo {author} {\bibfnamefont {Y.}~\bibnamefont
  {Couder}}, \bibinfo {author} {\bibfnamefont {L.}~\bibnamefont {Pauchard}},
  \bibinfo {author} {\bibfnamefont {C.}~\bibnamefont {Allain}}, \bibinfo
  {author} {\bibfnamefont {M.}~\bibnamefont {Adda-Bedia}}, \ and\ \bibinfo
  {author} {\bibfnamefont {S.}~\bibnamefont {Douady}},\ }\bibfield  {title}
  {\enquote {\bibinfo {title} {{The leaf venation as formed in a tensorial
  field}},}\ }\href {\doibase 10.1140/epjb/e2002-00211-1} {\bibfield  {journal}
  {\bibinfo  {journal} {The European Physical Journal B}\ }\textbf {\bibinfo
  {volume} {28}},\ \bibinfo {pages} {135--138} (\bibinfo {year}
  {2002})}\BibitemShut {NoStop}%
\bibitem [{\citenamefont {Mizuno}\ \emph {et~al.}(2007)\citenamefont {Mizuno},
  \citenamefont {Tardin}, \citenamefont {Schmidt},\ and\ \citenamefont
  {MacKintosh}}]{Mizuno2007}%
  \BibitemOpen
  \bibfield  {author} {\bibinfo {author} {\bibfnamefont {Daisuke}\ \bibnamefont
  {Mizuno}}, \bibinfo {author} {\bibfnamefont {Catherine}\ \bibnamefont
  {Tardin}}, \bibinfo {author} {\bibfnamefont {C.~F.}\ \bibnamefont {Schmidt}},
  \ and\ \bibinfo {author} {\bibfnamefont {F.~C.}\ \bibnamefont {MacKintosh}},\
  }\bibfield  {title} {\enquote {\bibinfo {title} {{Nonequilibrium Mechanics of
  Active Cytoskeletal Networks}},}\ }\href {\doibase 10.1126/science.1134404}
  {\bibfield  {journal} {\bibinfo  {journal} {Science}\ }\textbf {\bibinfo
  {volume} {315}},\ \bibinfo {pages} {370--373} (\bibinfo {year}
  {2007})}\BibitemShut {NoStop}%
\bibitem [{\citenamefont {Ronceray}\ \emph {et~al.}(2016)\citenamefont
  {Ronceray}, \citenamefont {Broedersz},\ and\ \citenamefont
  {Lenz}}]{Ronceray2016}%
  \BibitemOpen
  \bibfield  {author} {\bibinfo {author} {\bibfnamefont {Pierre}\ \bibnamefont
  {Ronceray}}, \bibinfo {author} {\bibfnamefont {Chase~P.}\ \bibnamefont
  {Broedersz}}, \ and\ \bibinfo {author} {\bibfnamefont {Martin}\ \bibnamefont
  {Lenz}},\ }\bibfield  {title} {\enquote {\bibinfo {title} {{Fiber networks
  amplify active stress}},}\ }\href {\doibase 10.1073/pnas.1514208113}
  {\bibfield  {journal} {\bibinfo  {journal} {Proceedings of the National
  Academy of Sciences}\ }\textbf {\bibinfo {volume} {113}},\ \bibinfo {pages}
  {2827--2832} (\bibinfo {year} {2016})}\BibitemShut {NoStop}%
\bibitem [{\citenamefont {Noll}\ \emph {et~al.}(2017)\citenamefont {Noll},
  \citenamefont {Mani}, \citenamefont {Heemskerk}, \citenamefont {Streichan},\
  and\ \citenamefont {Shraiman}}]{Noll2017}%
  \BibitemOpen
  \bibfield  {author} {\bibinfo {author} {\bibfnamefont {Nicholas}\
  \bibnamefont {Noll}}, \bibinfo {author} {\bibfnamefont {Madhav}\ \bibnamefont
  {Mani}}, \bibinfo {author} {\bibfnamefont {Idse}\ \bibnamefont {Heemskerk}},
  \bibinfo {author} {\bibfnamefont {Sebastian~J.}\ \bibnamefont {Streichan}}, \
  and\ \bibinfo {author} {\bibfnamefont {Boris~I.}\ \bibnamefont {Shraiman}},\
  }\bibfield  {title} {\enquote {\bibinfo {title} {Active tension network model
  suggests an exotic mechanical state realized in epithelial tissues},}\ }\href
  {\doibase 10.1038/nphys4219} {\bibfield  {journal} {\bibinfo  {journal}
  {Nature Physics}\ }\textbf {\bibinfo {volume} {13}},\ \bibinfo {pages}
  {1221--1226} (\bibinfo {year} {2017})}\BibitemShut {NoStop}%
\bibitem [{\citenamefont {Rocks}\ \emph {et~al.}(2017)\citenamefont {Rocks},
  \citenamefont {Pashine}, \citenamefont {Bischofberger}, \citenamefont
  {Goodrich}, \citenamefont {Liu},\ and\ \citenamefont {Nagel}}]{Rocks2017}%
  \BibitemOpen
  \bibfield  {author} {\bibinfo {author} {\bibfnamefont {Jason~W}\ \bibnamefont
  {Rocks}}, \bibinfo {author} {\bibfnamefont {Nidhi}\ \bibnamefont {Pashine}},
  \bibinfo {author} {\bibfnamefont {Irmgard}\ \bibnamefont {Bischofberger}},
  \bibinfo {author} {\bibfnamefont {Carl~P}\ \bibnamefont {Goodrich}}, \bibinfo
  {author} {\bibfnamefont {Andrea~J}\ \bibnamefont {Liu}}, \ and\ \bibinfo
  {author} {\bibfnamefont {Sidney~R}\ \bibnamefont {Nagel}},\ }\bibfield
  {title} {\enquote {\bibinfo {title} {{Designing allostery-inspired response
  in mechanical networks}},}\ }\href {\doibase 10.1073/pnas.1612139114}
  {\bibfield  {journal} {\bibinfo  {journal} {Proceedings of the National
  Academy of Sciences}\ }\textbf {\bibinfo {volume} {114}},\ \bibinfo {pages}
  {2520--2525} (\bibinfo {year} {2017})}\BibitemShut {NoStop}%
\bibitem [{\citenamefont {Kim}\ \emph {et~al.}(2019)\citenamefont {Kim},
  \citenamefont {Lu}, \citenamefont {Strogatz},\ and\ \citenamefont
  {Bassett}}]{Kim2019}%
  \BibitemOpen
  \bibfield  {author} {\bibinfo {author} {\bibfnamefont {Jason~Z.}\
  \bibnamefont {Kim}}, \bibinfo {author} {\bibfnamefont {Zhixin}\ \bibnamefont
  {Lu}}, \bibinfo {author} {\bibfnamefont {Steven~H.}\ \bibnamefont
  {Strogatz}}, \ and\ \bibinfo {author} {\bibfnamefont {Danielle~S.}\
  \bibnamefont {Bassett}},\ }\bibfield  {title} {\enquote {\bibinfo {title}
  {{Conformational control of mechanical networks}},}\ }\href {\doibase
  10.1038/s41567-019-0475-y} {\bibfield  {journal} {\bibinfo  {journal} {Nature
  Physics}\ }\textbf {\bibinfo {volume} {15}},\ \bibinfo {pages} {714--720}
  (\bibinfo {year} {2019})}\BibitemShut {NoStop}%
\bibitem [{\citenamefont {do~Carmo}(1988)}]{DoCarmo1988}%
  \BibitemOpen
  \bibfield  {author} {\bibinfo {author} {\bibfnamefont {Manfredo~P.}\
  \bibnamefont {do~Carmo}},\ }\href@noop {} {\emph {\bibinfo {title}
  {{Differential Geometry of Curves and Surfaces}}}}\ (\bibinfo  {publisher}
  {Prentice-Hall},\ \bibinfo {address} {London},\ \bibinfo {year}
  {1988})\BibitemShut {NoStop}%
\bibitem [{\citenamefont {Lubensky}\ \emph {et~al.}(2015)\citenamefont
  {Lubensky}, \citenamefont {Kane}, \citenamefont {Mao}, \citenamefont
  {Souslov},\ and\ \citenamefont {Sun}}]{Lubensky2015}%
  \BibitemOpen
  \bibfield  {author} {\bibinfo {author} {\bibfnamefont {T.~C.}\ \bibnamefont
  {Lubensky}}, \bibinfo {author} {\bibfnamefont {C.~L.}\ \bibnamefont {Kane}},
  \bibinfo {author} {\bibfnamefont {Xiaoming}\ \bibnamefont {Mao}}, \bibinfo
  {author} {\bibfnamefont {Anton}\ \bibnamefont {Souslov}}, \ and\ \bibinfo
  {author} {\bibfnamefont {Kai}\ \bibnamefont {Sun}},\ }\bibfield  {title}
  {\enquote {\bibinfo {title} {{Phonons and elasticity in critically
  coordinated lattices}},}\ }\href {\doibase 10.1088/0034-4885/78/7/073901}
  {\bibfield  {journal} {\bibinfo  {journal} {Reports on Progress in Physics}\
  }\textbf {\bibinfo {volume} {78}},\ \bibinfo {pages} {073901} (\bibinfo
  {year} {2015})}\BibitemShut {NoStop}%
\bibitem [{\citenamefont {Bruyneel}\ and\ \citenamefont
  {Duysinx}(2005)}]{Bruyneel2005}%
  \BibitemOpen
  \bibfield  {author} {\bibinfo {author} {\bibfnamefont {M.}~\bibnamefont
  {Bruyneel}}\ and\ \bibinfo {author} {\bibfnamefont {P.}~\bibnamefont
  {Duysinx}},\ }\bibfield  {title} {\enquote {\bibinfo {title} {{Note on
  topology optimization of continuum structures including self-weight}},}\
  }\href {\doibase 10.1007/s00158-004-0484-y} {\bibfield  {journal} {\bibinfo
  {journal} {Structural and Multidisciplinary Optimization}\ }\textbf {\bibinfo
  {volume} {29}},\ \bibinfo {pages} {245--256} (\bibinfo {year}
  {2005})}\BibitemShut {NoStop}%
\end{thebibliography}%

\newpage

\onecolumngrid
\beginsupplement

\newpage

\begin{center}
\large\textbf{Supplemental Material}
\end{center}

\section{Small-angle approximation}
\begin{figure}[h]
    \centering
    \includegraphics[width=0.3\columnwidth]{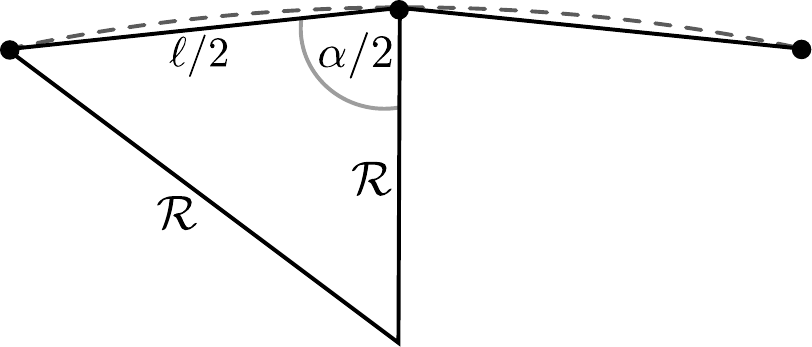}
    \caption{Approximating a weakly bent beam of length $\ell$
    (dashed line) by
    discrete elements (nodes are black circles). The bending
    angle between the discrete elements is $\alpha$,
    each element has length $\ell/2$, the beam's radius of
    curvature is $\mathcal{R}$.}
    \label{fig:small-angle}
\end{figure}
We now justify the small-angle approximation for the radius
of curvature of a weakly bent beam using the
setup shown in Fig.~\ref{fig:small-angle}. The law of cosines
in the shown triangle leads to
\begin{align*}
    \cos \frac{\alpha}{2} = \frac{1}{4} \frac{\ell}{\mathcal{R}}.
\end{align*}
Using the fact that $\alpha = \pi - \varepsilon$ for a small
angle $\varepsilon$, we simplify the left hand side as
\begin{align*}
    \cos \frac{\alpha}{2} = \sin \frac{\varepsilon}{2}
    \approx \frac{1}{2} \sin \varepsilon = \frac{1}{2} \sin \alpha,
\end{align*}
where we used that $\sin\varepsilon = 2\sin (\varepsilon/2)
\cos(\varepsilon/2)$ and $\cos(\varepsilon/2) \approx 1$.
We then find
\begin{align*}
    \frac{1}{\mathcal{R}^2} = \frac{4}{\ell^2}\sin^2\alpha,
\end{align*}
from which Eq.~(1) follows.

\section{Minimization over the rotational degrees of freedom}
We first show that Eq.~(2) reduces to the
correct elastic energy upon minimizing over the orientation
of the local reference frame.
We write the unit vectors defining the edges
in polar coordinates as $\mathbf{b}_1=(\sin\theta_1,\cos\theta_1,0)$,
$\mathbf{b}_2=(\sin\theta_2,\cos\theta_2,0)$.
The reference frame is chosen as
$\mathbf{e}_1 = (1,0,0)$ and $\mathbf{e}_2=(-1,0,0)$.
Then, the rotated reference frame can be expressed as $R\,\mathbf{e}_1=(\sin\phi,\cos\phi,0)$,
and $R\,\mathbf{e}_2 = -R\,\mathbf{e}_1$,
where $\phi$ is the angle of rotation in the $x$--$y$ plane.
With this, the elastic energy Eq.~(2) becomes,
\begin{align*}
    V = \frac{1}{2}\tilde{\kappa}_b\sin^2(\phi - \theta_1) + \frac{1}{2}\tilde{\kappa}_b\sin^2(\pi + \phi - \theta_2).
\end{align*}
In the limit of small angles, the minimizer
of $V$ with respect to $\phi$ is $\phi^* \approx (\theta_1 + \theta_2-\pi)/2$. Plugging
this back in we obtain
\begin{align*}
    V \approx \frac{1}{4}\tilde{\kappa}_b\,\sin^2(\pi + \theta_1 - \theta_2) =
    \frac{1}{4}\tilde{\kappa}_b \sin^2\alpha,
\end{align*}
which agrees with Eq.~(1) upon identifying
$\tilde{\kappa}_b = 2\kappa_b$.

\section{DBN bending energy}
We now derive the DBN bending energy
Eq.~(4) from Eq.~(3).
We minimize the elastic energy Eq.~(3) over
the linearized rotation matrix which is parametrized by
a vector $\mathbf{n}_i$ and acts as $R_i\,\mathbf{a} \approx
\mathbf{a} + \mathbf{n}_i\times\mathbf{a}$ on a
vector $\mathbf{a}$.
%
We write the position of each node as $\mathbf{x}_i =
\mathbf{x}_i^{(0)} + \mathbf{u}_i$, where $\mathbf{x}_i^{(0)}$
is the equilibrium position and $\mathbf{u}_i$
is a small displacement. To linear order, the unit vector
along an edge $b=(ij)$ can be expanded as $\mathbf{b} \approx
\mathbf{e}_b + J_b (\mathbf{u}_j - \mathbf{u}_i)$, where
the Jacobian encodes the double cross product $J_b \mathbf{a}
= -\frac{1}{\ell_b} \mathbf{e}_b \times (\mathbf{e}_b \times \mathbf{a})$ with
the equilibrium length $\ell_b$ and the equilibrium unit
vector $\mathbf{e}_b=(\mathbf{x}_j^{(0)} - \mathbf{x}_i^{(0)})/\|\mathbf{x}_j^{(0)} - \mathbf{x}_i^{(0)}\|$.
With this, we can expand
\begin{align*}
    \| (R_i\, \mathbf{e}_b) \times \mathbf{b} \|^2 &\approx
    \| (\mathbf{e}_b + J_b(\mathbf{u}_j - \mathbf{u}_i))
    \times (\mathbf{e}_b + \mathbf{n}_i\times \mathbf{e}_b) \|^2 \\
    &= \| D_b \mathbf{u} - C_b\mathbf{n}_i \|^2,
\end{align*}
where we neglected non-linear terms in $\mathbf{u}$ and
$\mathbf{n}_i$.
Here, the matrix $D_b$ acts as $D_b \mathbf{u}
= \frac{1}{\ell_b} \mathbf{e}_b \times (\mathbf{u}_j
-\mathbf{u}_i)$, $C_b = \mathbb{1} -
\mathbf{e}_b\mathbf{e}_b^\top$, and the
$3N$-dimensional vector $\mathbf{u}$
contains the displacements of the $N$ nodes.
At each node $i$, the linearized elastic energy is then
\begin{align}
    V_i
    &= \frac{1}{2} \sum_{b\in B_i} \kappa_b \| D_b\mathbf{u} - C_b\mathbf{n}_i \|^2. \label{eq:linearized-Vi}
\end{align}
Taking the gradient of $V_i$ with respect to $\mathbf{n}_i$
and setting it to zero using $C_b^\top C_b = C_b$ and $C_b^\top D_b = D_b$ we obtain
    $C_i \mathbf{n}_i = D_i\mathbf{u}$,
where $C_i = \sum_{b\in B_i} \kappa_b C_b$ and $D_i = \sum_{b\in B_i} \kappa_b
D_b$.
Formally solving this linear equation for $\mathbf{n}_i$,
plugging the result into \eqref{eq:linearized-Vi},
and summing over all nodes $i$, we arrive at
Eq.~(4) with
\begin{align}
    H_\text{eq} =
\sum_i\sum_{b\in B_i} \kappa_b D_b^\top D_b \label{eq:Heq}
\end{align}
and
\begin{align}
    H_\text{or} = \sum_i D_i^\top C_i^{-1} D_i. \label{eq:Hor}
\end{align}

We note that although we derived
Eq.~(4) from Eq.~(2),
which models only the lowest bending mode, larger DBNs can
naturally model higher modes as well if they contain many connected nodes
arranged in a line, providing a fine discretization of a continuum beam.

\subsection{Planar networks}
For planar, inextensible networks, it can be shown that only the
$z$ component of the displacements $\mathbf{u}$ is nonzero (Section~\ref{sect:planar}). With this, \eqref{eq:Heq}
corresponds to the weighted network Laplacian,
\begin{align*}
    \frac{1}{2} \mathbf{u}^\top H_\mathrm{eq} \mathbf{u}
    &= \frac{1}{2}\sum_i\sum_{b\in B_i} \frac{\kappa_b}{\ell_b^2}
    \|\mathbf{e}_b \times (\mathbf{u}_j - \mathbf{u}_i)\|^2 \\
    &= \sum_{i,j} \frac{\kappa_b}{\ell_b^2} (u_{z,j} - u_{z,i})^2.
\end{align*}
This expression depends only on the weighted topology of the elastic
network and not on the geometry at all.
In contrast, \eqref{eq:Hor} can not be written in a purely
topological way and encodes the geometry and weights in a
nontrivial way. We find in terms of the displacements,
\begin{align*}
    \frac{1}{2} \mathbf{u}^\top H_\mathrm{or} \mathbf{u}
    &= \frac{1}{2}\sum_i \left( \sum_{b\in B_i} \frac{\kappa_b}{\ell_b}
    \mathbf{e}_b \times (\mathbf{u}_j - \mathbf{u}_i)\right)^\top
    C_i^{-1} \left( \sum_{b\in B_i} \frac{\kappa_b}{\ell_b}
    \mathbf{e}_b \times (\mathbf{u}_j - \mathbf{u}_i)\right) \\
    &= \frac{1}{2}\sum_i \left( \sum_{b\in B_i} \frac{\kappa_b}{\ell_b}
    (u_{z,j} - u_{z,i})\,
    \mathbf{e}_b^\perp \right)^\top
    C_i^{-1} \left( \sum_{b\in B_i} \frac{\kappa_b}{\ell_b}
    (u_{z,j} - u_{z,i})\,
    \mathbf{e}_b^\perp\right),
\end{align*}
where $\mathbf{e}_b$ is rotated by $\pi/2$ in the $x$--$y$ plane
into $\mathbf{e}_b^\perp$.
The matrix $C_i^{-1} = \left(\sum_{b\in B_i} \kappa_b\, (\mathbb{1} - \mathbf{e}_b \mathbf{e}_b^\top)\right)^{-1}$ depends on both stiffnesses
and local geometry at the node $i$.

\section{Nodal force balance}
\label{sect:force-balance} %
We now derive the nodal force balance from \eqref{eq:linearized-Vi}.
Rewriting in terms of three-dimensional vectors and making the edges
$b=(ij)$ explicit the total network energy $V=\sum_i V_i$ reads
\begin{align*}
    V
    &= \frac{1}{2} \sum_{i,j} \kappa_{ij} \| C_{ij}\mathbf{n}_i -
    \ell_{ij}^{-1} \mathbf{e}_{ij} \times (\mathbf{u}_j - \mathbf{u}_i) \|^2.
\end{align*}
Using $\partial V/\partial \mathbf{n}_i^\top = 0$, the net force on node $i$ is
\begin{align*}
    \mathbf{F}_i &= -\frac{\partial V}{\partial \mathbf{u}_i^\top} = \sum_j (\mathbf{F}_{ij} - \mathbf{F}_{ji}),
\end{align*}
where we used that each nodal displacement $\mathbf{u}_i$ appears in
$V_i$ and in all $V_j$ that are connected to node $i$. The forces are
\begin{align*}
    \mathbf{F}_{ij} = -\frac{\kappa_{ij}}{\ell_{ij}} \mathbf{e}_{ij} \times
        \left( C_{ij}\mathbf{n}_i -
    \ell_{ij}^{-1} \mathbf{e}_{ij} \times (\mathbf{u}_j - \mathbf{u}_i) \right).
\end{align*}
Here, $\kappa_{ij} = \kappa_{ji}$, $\ell_{ij} = \ell_{ji}$, and $C_{ij} = C_{ji}$. Using the definition of $C_{ij} = \mathbb{1} - \mathbf{e}_{ij}
\mathbf{e}_{ij}^\top$, the magnitudes are,
\begin{align}
    \|\mathbf{F}_{ij}\|^2 &= \frac{\kappa_{ij}^2}{\ell_{ij}^2}
    \| C_{ij}\mathbf{n}_i -
    \ell_{ij}^{-1} \mathbf{e}_{ij} \times (\mathbf{u}_j - \mathbf{u}_i) \|^2. \label{eq:nodal-force-norm}
\end{align}

\section{Constrained Optimization}
\label{sect:optimization}
We adapt the global approach outlined in Ref.~\cite{Katifori2010b}.
The Lagrangian corresponding to the constrained minimization problem
is
\begin{align*}
    \mathcal{L}(\{\kappa_b\}) = c(\{\kappa_b+\kappa_0\}) + \lambda \big(
    \sum_b \kappa_b^\gamma - K\big),
\end{align*}
where $c = \mathbf{f}^\top\mathbf{u}$ is the compliance and $\lambda$ is a Lagrange multiplier.
Taking the gradient with respect to $\kappa_b$ and
combining with \eqref{eq:nodal-force-norm}
leads to the scaling relation Eq.~(5).
We numerically solve for the $\kappa_b$ using the iteration
\begin{align}
    \tilde \kappa_b^{(n+1)} &= \left(
    -(\kappa_b^{(n)})^2 \frac{\partial
    c(\{\kappa_b^{(n)}+\kappa_0\})}{\partial \kappa_b}
    \right)^{1/(\gamma+1)}
    \nonumber \\
    \kappa_b^{(n+1)} &= \frac{\tilde \kappa_b^{(n+1)}}{\left(\sum_{b'}  (\tilde\kappa_{b'}^{(n+1)})^\gamma\right)^{1/\gamma}}, \label{eq:fixed-point}
\end{align}
where the second step fixes the Lagrange multiplier by enforcing the
constraint.
Combining \eqref{eq:fixed-point} with a variant of simulated
annealing leads to approximate global minimization.
At every $N_{\mathrm{therm}}$-th
step of the iteration \eqref{eq:fixed-point}, the $\{\kappa_b\}$ are
first thermalized by convolving with a Gaussian
kernel $G_{ab} \sim \exp(-d_{ab}^2/(2\sigma^2))$ where $d_{ab}$
is the Euclidean distance between edges $a$ and $b$ and where
the scale $\sigma$ is decreased after each thermalization.
Then, multiplicative noise $\sim \exp(s\, \xi)$, where $\xi$ is
normally distributed and $s\sim\mathcal{O}(1)$, is applied.
After a set number of thermalization steps, \eqref{eq:fixed-point}
is iterated until convergence.

\section{Metamaterials}
\begin{figure}
    \centering
    \includegraphics[width=.8\columnwidth]{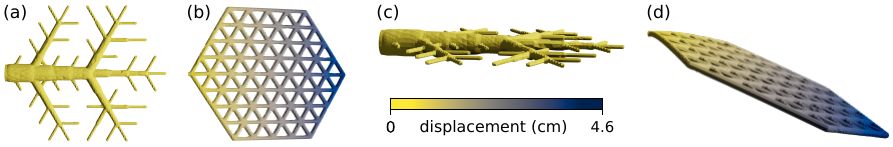}
    \caption{(a) 3D mesh used to manufacture the optimized network from Fig. 4 (a).
    (b) 3D mesh used to manufacture the uniform network from Fig. 4 (b).
    (c, d) FEM simulations of the meshes from (a,b) clamped at the left side
    with downward gravitational load.}
    \label{fig:fem}
\end{figure}
3D meshes [Fig.~\ref{fig:fem} (a,b)] were constructed from cylinders with spherical
end-caps, with cylinder radii taken from optimal and uniform DBN models.
The metamaterials were commercially manufactured from thermoplastic
polyurethane (Materialise nv, Leuven, Belgium).
Finite Element Method simulations [Fig.~\ref{fig:fem} (c,d)] were performed using the MATLAB 2018b PDE Toolbox (The MathWorks, Inc.,
Natick, MA) and are consistent with the experimental results shown in Fig.~4.
Material properties were Young's Modulus
$Y=85\,\mathrm{MPa}$, density $\rho=1100\,\mathrm{kg}\,\mathrm{m}^{-3}$,
Poisson's ratio $\nu=0.49$.

\section{Continuum limit}
\label{sect:helfrich}
Here we demonstrate that the bending energy Eq.~(4)
in the continuum limit of an initially flat, uniform sheet in equilibrium
is equivalent to the Helfrich free energy~\cite{Helfrich1973,Safran1999,Seung1988},
\begin{align}
    F = \int_A ( \kappa_1 H^2 + \kappa_2 K )\,  dA,
    \label{eq:helfrich}
\end{align}
where $H$ and $K$ are the surface's mean and Gaussian curvatures, respectively,
$\kappa_1$,$\kappa_2$ are elastic constants, and the integral is over the
surface of the sheet $A$.
We choose a triangular grid to model the
flat sheet in the $x$--$y$ plane and set all the bending constants $\kappa_b$ to unity. The inextensibility constraint
is then equivalent to only allowing
displacements in the $z$ direction, $\mathbf{u} = (0,0,\mathbf{u}_z^\top)^\top$, since all local in-plane displacements
are forbidden.
At each node, the unit vectors in the directions of the edges are
\begin{align*}
\mathbf{e}_1&=(1,0,0)^\top, &\mathbf{e}_2&=(1/2,\sqrt{3}/2,0)^\top, \\
\mathbf{e}_3&=(1/2,-\sqrt{3}/2,0)^\top, &\mathbf{e}_4 &= -\mathbf{e}_1, \\
\mathbf{e}_5 &= -\mathbf{e}_2, &\mathbf{e}_6 &= -\mathbf{e}_3,
\end{align*}
and
the matrix $C_i = \operatorname{diag}(3,3,6)$.
In the limit where the edge lengths $\ell$ tend to zero, the sheet's
displacements are approximated by a height function $u_z = h(x,y)$.
The expressions involving the matrices $D_b$ can then be written
as $D_b\mathbf{u} \approx \ell^{-1} \mathbf{e}_b \times (0, 0,
h(x + \ell\, (\mathbf{e}_b)_x, y + \ell\, (\mathbf{e}_b)_y) -h(x,y))^\top$.
Plugging this form into Eq.~(4), expanding
to lowest order in $\ell$ and summing over all vertices we find for the
total bending energy,
\begin{align}
    V \approx \frac{3}{16} \sum_i \left( 3(h_{xx} + h_{yy})^2
    -4 (h_{xx} h_{yy} - h_{xy}^2)\right) \ell^2.
    \label{eq:dbn-uniform}
\end{align}
Using the small-gradient expansions~\cite{DoCarmo1988}
of the mean curvature $H \approx h_{xx} + h_{yy}$ and the
Gaussian curvature $K \approx h_{xx} h_{yy} - h_{xy}^2$, and
the area element $dA \approx \ell^2 \sqrt{3}/2$ corresponding to
hexagons around each node, in the limit $\ell \to 0$ the sum
\eqref{eq:dbn-uniform} tends to
the integral \eqref{eq:helfrich} with the
elastic constants $\kappa_1=3\sqrt{3}/8$ and $\kappa_2=-\sqrt{3}/2$.


\section{Mechanical constraints}
\label{sect:planar}
\subsection{Edge inextensibility}
We now discuss the effect of inextensible edges to lowest order.
The length of each edge $b = (ij)$ can be written as
\begin{align*}
    \ell_b &= \sqrt{\left(\mathbf{x}_i^{(0)} - \mathbf{x}_j^{(0)} + \mathbf{u}_i - \mathbf{u}_j\right)^2} \\
        &\approx \ell^{(0)}_b + \mathbf{e}_b^\top (\mathbf{u}_i - \mathbf{u}_j),
\end{align*}
where $\mathbf{x}_i^{(0)}$ is the equilibrium position of node $i$,
$\mathbf{e}_b = (\mathbf{x}^{(0)}_i - \mathbf{x}_j^{(0)})/
\|\mathbf{x}^{(0)}_i - \mathbf{x}_j^{(0)}\|$ is the edge unit vector and
we expanded to linear order.
The inextensibility constraint is then equivalent to allowing only
displacements satisfying the constraint $\mathbf{e}_b^\top (\mathbf{u}_i - \mathbf{u}_j)=0$
for all edges $b$. This can be implemented by assembling all these constraints
into a matrix $Q$ acting on the vector of all displacements in the
network and demanding
\begin{align}
    0 = Q\mathbf{u} = \begin{pmatrix}
    Q_x \\
    Q_y \\
    0
    \end{pmatrix}
    \begin{pmatrix}
    \mathbf{u}_x \\
    \mathbf{u}_y \\
    \mathbf{u}_z
    \end{pmatrix}.\label{eq:constraint}
\end{align}
Here, $\mathbf{u}_{x,y,z}$ are the $x$, $y$, and $z$ components of the
displacements, and we used the fact that we consider planar networks such that all unit vectors
$\mathbf{e}_b$ lie in the $x$--$y$ plane.
The matrix $Q$ is also known as the \emph{compatibility
matrix}~\cite{Lubensky2015} of the
elastic network, and its nullspace $Q\mathbf{u} = 0$ encodes the allowed displacements
satisfying the inextensibility constraint.
Inspecting \eqref{eq:constraint}, we find that
all displacements in the $z$ direction (perpendicular to the
network) $\mathbf{u}_z$ are allowed.

Any remaining degrees of freedom $\mathbf{u}_{x,y}$ must then be in-plane.
Non-degenerate triangulations (including triangular grids as used in the main paper) possess no such degrees of freedom except for Euclidean motions
(overall rotations and translations): each
triangle
is rigid by itself, and adding another triangle to an already rigid
finite triangular grid can not introduce in-plane soft modes as long
as it is joined by one of its sides.
This induction step can be seen as follows.
Each new triangle contributes two new edges and one new node.
Two new constraints corresponding
to two new edges are introduced. Since all nodes except for the
new one are already rigid, they will remain so, and
their in-plane degrees of freedom are all $\mathbf{u}_{x,y;i} = 0$.
The new in-plane degree of freedom $\mathbf{u}_{x,y}^*$ must then
satisfy $\mathbf{e}_1^*{}^\top\mathbf{u}_{x,y}^* = 0$
and $\mathbf{e}_2^*{}^\top\mathbf{u}_{x,y}^* = 0$, where $\mathbf{e}_{1,2}^*$
are the unit vectors corresponding to the newly added edges.
We assumed the triangulation to be non-degenerate, meaning
that the new edges are not parallel. From this, $\mathbf{u}^*_{x,y} = 0$
immediately follows.

We conclude that to linear order,
the allowed displacements
for non-degenerate triangulations are the
$\mathbf{u}_z$ in the $z$ direction, perpendicular to the planar network, as well as Euclidean transformations
(overall rotations and translations).

\subsection{Lamina inextensibility}
In many biological networks such as leaves, not just the veins are inextensible,
but also the lamina itself.
Stretching of a triangulated sheet can be
modeled using springs between nearest neighbors~\cite{Seung1988,Witten2007}.
Because the edges in the DBNs considered here are
inextensible, this automatically models an inextensible lamina as
well.

\subsection{Numerical implementation}
For the numerical optimizations, we explicitly construct a constraint matrix
by taking the compatibility matrix $Q$ and adding rows corresponding
to (i) the removal of overall an twist degree of freedom along the $x$ axis
and (ii) clamping of the petiole.
We then numerically compute a matrix $\Phi$ of basis vectors of
its nullspace and proceed
to use the projected Hessian $H' = \Phi H \Phi^\top$, which encodes
only the allowed degrees of freedom.
We constrain the lengths of the triangular grid topology
without removing edges whose stiffness is set to zero
by the optimization algorithm,
such that the lamina always remains inextensible.

\subsection{Soft modes}
Even with the constraints as implemented above, in-plane soft modes are possible, for instance
if the underlying network is chosen to be non-triangular,
with hypostatic coordination number $z<4$. However, this is unphysical in the biological
systems we aim to model.
Out-of-plane soft modes are possible if the lamina stiffness $\kappa_0=0$ and a node
is not connected by any nonzero $\kappa_b$ to other nodes. Then the
constraints allow arbitrary displacements in $z$ direction that are no longer energetically
penalized. This case is not observed
in optimized networks, as it would lead to very large compliance (the load would be
parallel to the soft mode displacement).

\section{Optimization with self-loads}
While for many biological systems such as leaves uniform loads
$\mathbf{g}$ are a reasonable approximation~\cite{John2017}, the case
of self-loads (i.e., loads that depend on the edge stiffnesses) can be
considered as well.
In general, the mathematical structure of self-loaded optimization problems
changes significantly~\cite{Bruyneel2005}, making the numerical methods employed
in the main paper inappropriate. Here, we consider a simple model of
self-loads and derive an iterative scheme to solve
the associated KKT (Karush--Kuhn--Tucker) optimality equations.

\subsection{Numerical optimization with self-loads}
To include self-loads we write the compliance in the form
\begin{align*}
    c = \mathbf{f}^\top H \mathbf{f},
\end{align*}
where now both the Hessian $H=H(\{\kappa_e\})$ and the loads
$\mathbf{f} =\mathbf{f}(\{\kappa_e\})$ are functions of the stiffnesses.
The gradient of the compliance with respect to the stiffnesses $\kappa_e$ is then
\begin{align}
    \frac{\partial c}{\partial \kappa_e} = - \mathbf{f}^\top \frac{\partial H}{\partial \kappa_e} \mathbf{f}
    + 2\, \mathbf{f}^\top H \frac{\partial \mathbf{f}}{\partial \kappa_e}.
    \label{eq:grad-self-loads}
\end{align}
Using the method of Lagrange multipliers to include the cost constraint and
the inequality constraint $\kappa_e \geq 0$, we derive the KKT equations
\begin{align}
    \kappa_e \frac{\partial c}{\partial \kappa_e} + \lambda \gamma\, \kappa_e^\gamma &= 0 \label{eq:kkt} \\
    \sum_e \kappa_e^\gamma = K. \nonumber
\end{align}
We first solve for the Lagrange multiplier by summing over
\eqref{eq:kkt}, obtaining
\begin{align*}
    \lambda = \frac{1}{\gamma K}\sum_e \left(
    \kappa_e\mathbf{f}^\top \frac{\partial H}{\partial \kappa_e} \mathbf{f}
    - 2\kappa_e\, \mathbf{f}^\top H \frac{\partial \mathbf{f}}{\partial \kappa_e}\right).
\end{align*}
Next, we multiply \eqref{eq:kkt} by $\kappa_e$ and rearrange to
\begin{align*}
    \kappa_e^{\gamma+1} \left(
    \lambda\gamma + 2 \kappa_e^{1-\gamma} \mathbf{f}^\top H \frac{\partial \mathbf{f}}{\partial \kappa_e}\right)
    = \kappa_e^2\, \mathbf{f}^\top \frac{\partial H}{\partial \kappa_e} \mathbf{f}.
\end{align*}
Since the right-hand side of this equation is non-negative, the left-hand side is also.
Furthermore, the right-hand side
corresponds to the same expression
involving physical forces as in the
main paper.
We can rearrange the expression above into the self-consistency equation
\begin{align*}
    \kappa_e
    = \left(\frac{\kappa_e^2\, \mathbf{f}^\top \frac{\partial H}{\partial \kappa_e} \mathbf{f}}
    { \lambda\gamma + 2 \kappa_e^{1-\gamma} \mathbf{f}^\top H \frac{\partial \mathbf{f}}{\partial \kappa_e}}\right)^{\frac{1}{1+\gamma}}.
\end{align*}
From this we construct the iterative scheme
\begin{align*}
    \tilde{\kappa}_e^{(n+1)}
    &= \left.\left|\frac{\kappa_e^2\, \mathbf{f}^\top \frac{\partial H}{\partial \kappa_e} \mathbf{f}}
    { \lambda\gamma + 2 \kappa_e^{1-\gamma} \mathbf{f}^\top H \frac{\partial \mathbf{f}}{\partial \kappa_e}}\right|^{\frac{1}{1+\gamma}}\right.^{(n)} \\
    \kappa_e^{(n+1)} &= K^{1/\gamma}\frac{\tilde{\kappa}_e^{(n)}}{\left(\sum_f (\tilde{\kappa}_f^{(n)})^\gamma\right)^{1/\gamma}},
\end{align*}
where the absolute value is taken to avoid negative values that may appear due to numerical issues at small values of $\kappa_e$. The constraint is explicitly enforced at each step to prevent
numerical constraint drifting.

\subsection{A simple model for mass self-loads}
We now construct a simple model for self-loads based on vein mass.
We assume that the mass of each vein can be modeled as
\begin{align*}
    m_e = a\, \kappa_e^\alpha,
\end{align*}
where $a$ is a constant of proportionality and $\alpha$ is a parameter. We expect the biologically relevant regime to be close to $\alpha=1/2$ (solid cylindrical beams).
In this case, $a = \rho \ell^{3/2} \sqrt{\pi/Y}$.
With this, we write the nondimensional perpendicular load at each node $i$ as
\begin{align*}
    f_{z,i} = (1 - \beta)\,\bar{f} + \frac{\beta}{N}\,\frac{1}{N_n} \sum_{(e,i)}\kappa_e^\alpha,
\end{align*}
where $\bar{f}$ is the dimensionless load due to the lamina, $\beta$ controls the overall proportion of vein load and lamina load, and the sum is over all $N_n$
edges $e$ neighboring node $i$. The additional factor of $1/N$, where $N$
is the number of nodes in the network, serves to bring the two terms to
roughly the same scale. Any overall constants of proportionality are absorbed into the total cost $K$.

\subsection{Results}
Since the biological regime is expected to be near $\alpha=1/2$,
the following, we specialize the case $\alpha=1/2$ and $\gamma=1/2$.
We also set $\bar{f} = 1/N$, where $N$ is the number of nodes in the network and look at the biologically relevant regime
where $0 \leq \beta \leq 0.5$. We note that the fraction
of vein mass to total mass $\mu$ can only be evaluated
a posteriori, but is generally observed to be close to $\beta$ with
the normalizations chosen above.

\begin{figure}
    \centering
    \includegraphics[width=\textwidth]{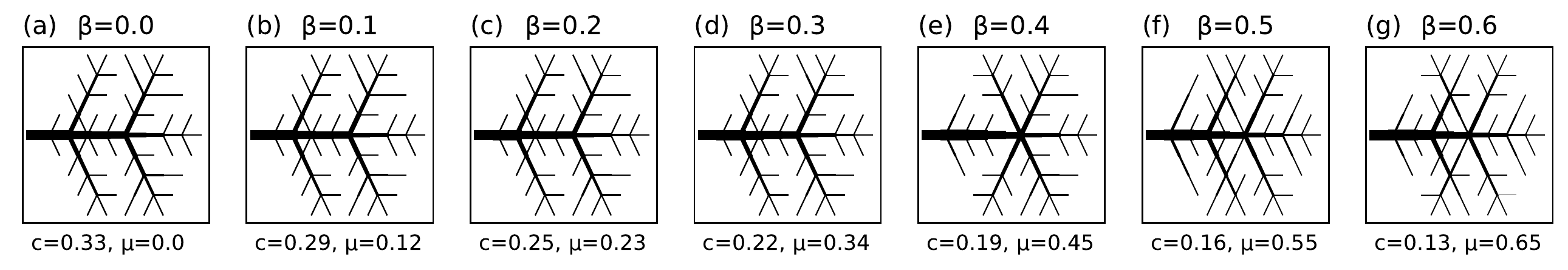}
    \caption{Optimal network topologies including self-loads with parameter $\beta$
    describing the ratio between uniform lamina load and vein load and
    optimal compliance $c$. The final proportion
    $\mu$ of vein weight to total weight is calculated a posteriori.
    For values of $\mu\lesssim 0.34$, the optimal network topology is identical.
    Thus, no self-loads ($\beta=0$) are a reasonable approximation of real biological
    networks.}
    \label{fig:self-loads}
\end{figure}

We find that up to a value of $\mu \approx 0.35$, the optimal
networks with self-loads have the same topology (but not compliance value
or exact numerical value of the $\kappa_e$) as those with $\beta=0$ (no
self loads), such that neglecting self-loads appears to be a reasonable
approximation (Fig.~\ref{fig:self-loads}).
Numerical experiments also indicate that in this regime, the two algorithms
with and without self-loads converge to the same final network topology
from identical initial conditions if no simulated annealing is used.
For larger values of $0.4 \leq \beta \leq 0.6$ the optimal topologies start to differ, but
not in a drastic way.
In the less biologically relevant regime $\beta \gg 0.6$,
the numerical scheme suffers
from instabilities and often does not converge.

\section{Scaling of the phase space of optimal DBNs}
\begin{figure}
    \centering
    \includegraphics[width=\textwidth]{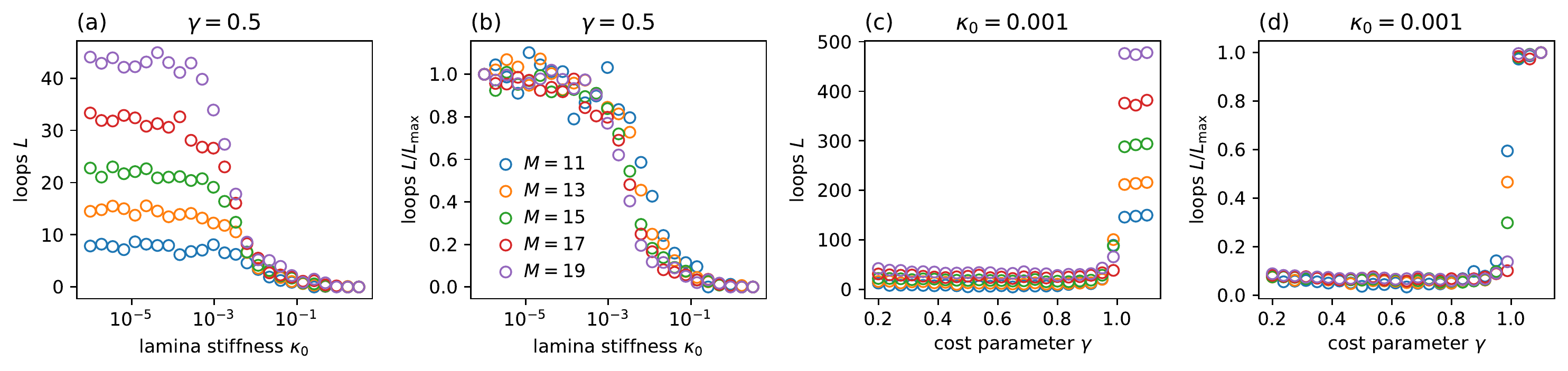}
    \caption{Scaling of the number of loops. (a) Number of loops at slice through the phase
    space at $\gamma = 0.5$. (b) Number of loops normalized by the maximum at
    slice through the phase
    space at $\gamma = 0.5$.
    (c) Number of loops at slice through the phase
    space at $\kappa_0 = 0.001$. (d) Number of loops normalized by the maximum at
    slice through the phase
    space at $\kappa_0 = 0.001$.}
    \label{fig:loops}
\end{figure}

\begin{figure}
    \centering
    \includegraphics[width=\textwidth]{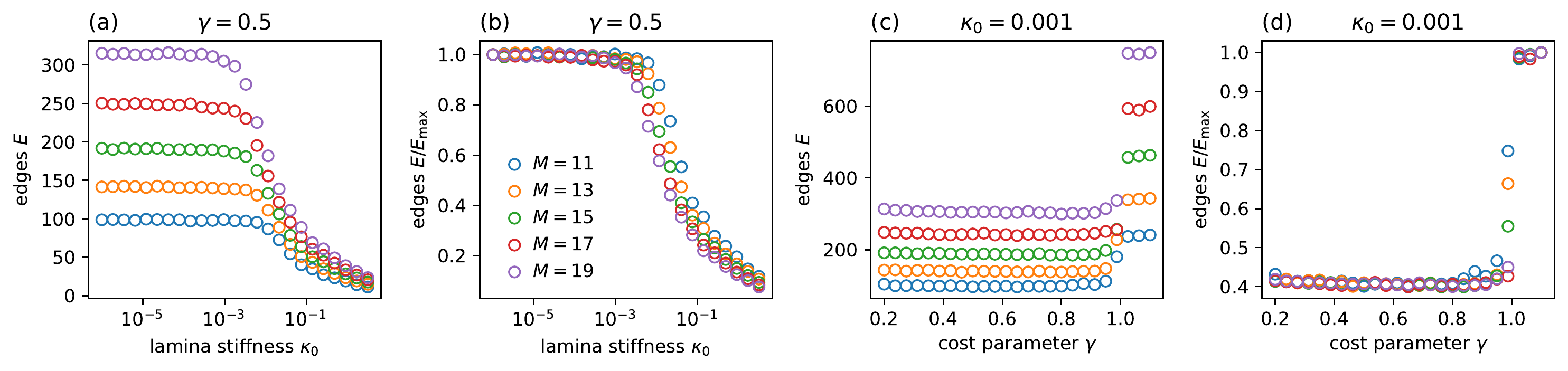}
    \caption{Scaling of the number of nonzero edges. (a) Number of edges at slice through the phase
    space at $\gamma = 0.5$. (b) Number of edges normalized by the maximum at
    slice through the phase
    space at $\gamma = 0.5$.
    (c) Number of edges at slice through the phase
    space at $\kappa_0 = 0.001$. (d) Number of edges normalized by the maximum at
    slice through the phase
    space at $\kappa_0 = 0.001$.}
    \label{fig:edges}
\end{figure}

\begin{figure}
    \centering
    \includegraphics[width=\textwidth]{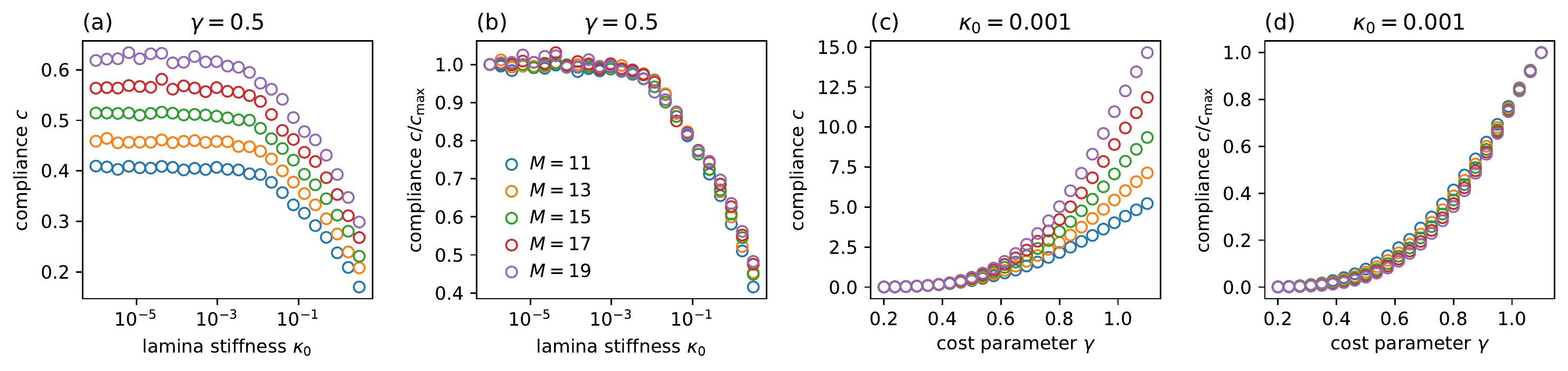}
    \caption{Scaling of the compliance. (a) Compliance at slice through the phase
    space at $\gamma = 0.5$. (b) Compliance normalized by the compliance
    of a uniform network with identical cost at slice through the phase
    space at $\gamma = 0.5$.
    (c) Compliance at slice through the phase
    space at $\kappa_0 = 0.001$. (d) Compliance normalized by the compliance
    of a uniform network with identical cost at slice through the phase
    space at $\kappa_0 = 0.001$.}
    \label{fig:compliance}
\end{figure}

Here we present a size scaling analysis of the topological phase
space shown in Fig.~3 of the main paper. While the phase space there was
computed for networks with 92 nodes, here we show slices through the phase
space for larger networks. We consider slices at $\gamma=0.5$ and $\kappa_0=10^{-3}$
and parametrize the networks by the linear number of nodes $M$ along the
midrib. For the triangular networks we consider, the total number of nodes
$N\sim \mathcal{O}(M^2)$.
The scaling of the number of loops is shown in Fig.~\ref{fig:loops},
the scaling of the number of nonzero edges is shown in Fig.~\ref{fig:edges},
and the scaling of the compliance is shown in Fig.~\ref{fig:compliance}.
We estimate the number of nonzero edges by thresholding the results of
the optimization at $\kappa_b = 10^{-8}$ and considering all edges with
smaller bending stiffness as absent.
Similarly, we estimate the number of nodes by computing the weighted degree
$d_i = (1/n) \sum_j \kappa_{ij}$ of each node $i$ in the original triangular network
with $n$ neighbors, and again count nodes with $d_i < 10^{-8}$ as absent.
Each data point in the aforementioned figures
shows an average over at least 10 optimizations.
All curves for different network sizes collapse after rescaling,
suggesting that the phase space shown in Fig.~3 of the main
paper is robust as network size is varied.

\section{Three-dimensional optimal DBNs}
\begin{figure}
    \centering
    \includegraphics[width=\textwidth]{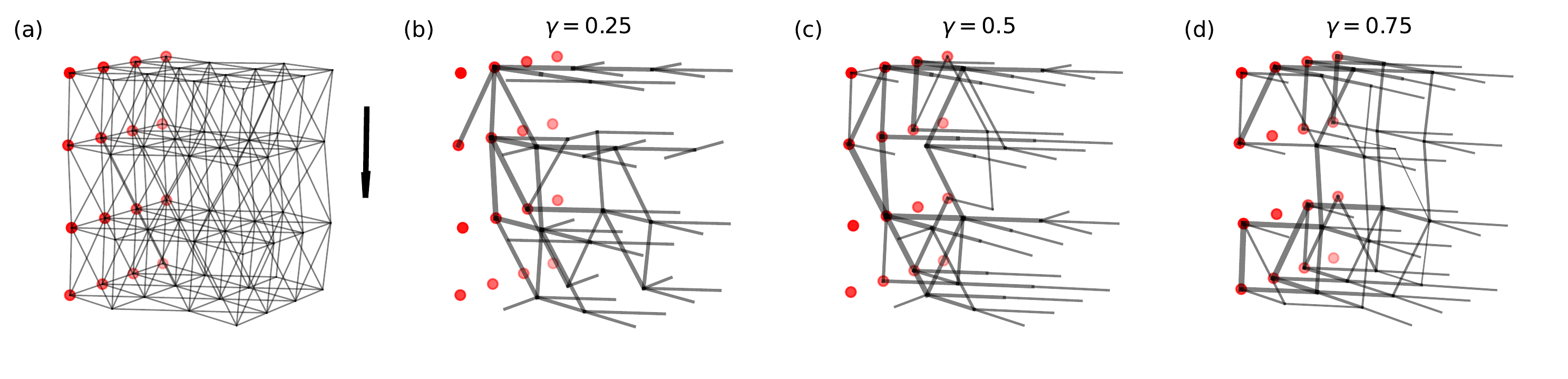}
    \caption{Three-dimensional optimal DBNs fixed
    at one side.
    (a) Tetrahedral base network with fixed
    nodes indicated in red. The uniform load $\mathbf{f}$
    is shown as a black arrow.
    (b--d) Optimal networks obtained using
    simulated annealing with cost parameters
    $\gamma=0.25, 0.5, 0.75$ and $\kappa_0=10^{-4}$. Line widths are
    proportional to $\kappa_b^{\gamma/2}$.}
    \label{fig:3d}
\end{figure}
Here we show that the DBN model introduced in the main paper
can be used to model fully three-dimensional networks of connected bending
beams as well.
As the base topology, we take a three-dimensional tetrahedral network
[Fig.~\ref{fig:3d}~(a)]. Since such a network is perfectly
rigid under the inextensibility constraint, for
the purposes of this proof of concept, we remove the constraint.
We note that for realistic applications, it would be necessary to introduce
a stretching energy including a relationship between stretching and
bending stiffnesses of each beam.
Optimal networks fixed at one side and under
uniform perpendicular load show similar features
as sheet-like DBNs [Fig.~\ref{fig:3d} (b--d)].

\section{Topological comparison to real leaf networks}
In this section, we compare the topology of optimized DBNs to that
of real leaf networks. In Ref.~\cite{Blonder2020}, Blonder~\emph{et~al.}
introduced the Minimum Spanning Tree (MST) ratio as an easily computable topological metric
to characterize leaf networks over many scales.
For a generic weighted network embedded in space, the MST ratio is defined as
\begin{align}
    \text{MST ratio} = \frac{\text{length of all edges in a minimum spanning tree}}{\text{total length of all edges}}.
\end{align}
To calculate the MST, we choose the inverse edge diameter as the weight to
preferentially incorporate large veins and then employ Kruskal's algorithm.

\begin{figure}
    \centering
    \includegraphics[width=\textwidth]{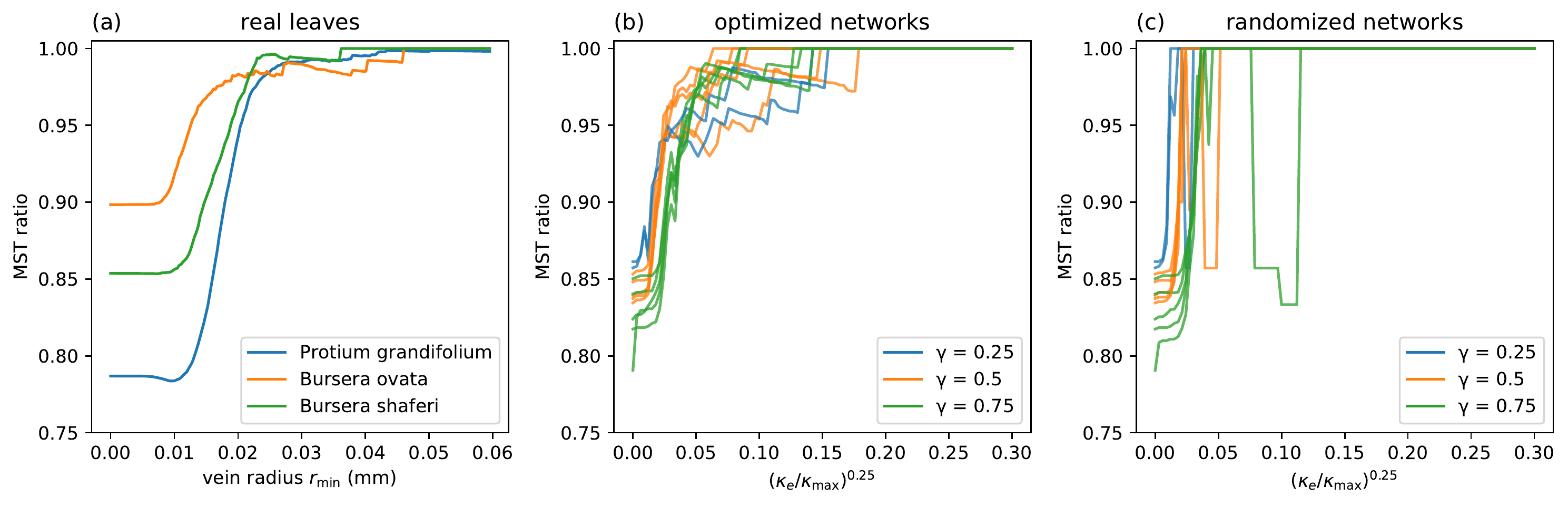}
    \caption{Topology of real leaf networks and optimized DBNs using the
    multi-scale MST ratio. (a) MST ratios for the three real leaf networks shown in
    Fig.~\ref{fig:leaves}. MST ratios are calculated for the network where
    all veins with radius less than $r_\mathrm{min}$ are discarded. (b) MST ratios
    for optimized annealed DBNs with $N=722$ nodes and $E=861$ edges and values of $\gamma\in\{0.25, 0.5, 0.75\}$. The value $(\kappa_e/\kappa_\mathrm{max})^{1/4}$ was used as a proxy for vein radius,
    corresponding to solid cylindrical beams. Here, $\kappa_\mathrm{max}$ is the maximum
    value of $\kappa_e$ over the entire network. (c) MST ratio curves for the same networks as in
    panel (b), but with the $\kappa_e$ randomly shuffled.}
    \label{fig:leaf-topology}
\end{figure}

Since leaf networks exhibit different structure at different scales, the MST ratio is calculated not just for
the entire network, but also for
pruned networks where all edges below a certain radius $r_\mathrm{min}$ are discarded. The
resulting values are plotted as a function of $r_\mathrm{min}$ to obtain a
graph characterizing the topology of the network.
We calculated this measure for three leaf networks from the data set of
Ref.~\cite{Ronellenfitsch2015b} and found similar characteristic curves as Blonder~\emph{et~al.}
[Fig.~\ref{fig:leaf-topology}~(a) and Fig.~\ref{fig:leaves}].
Generally, for small $r_\mathrm{min}$ the MST ratio is approximately constant, indicating that small
veins are approximately equally distributed between loops and branches.
After some critical radius, the MST ratio steeply increases as more loops than branches are removed.
As the larger scales of the network are reached, the MST ratio is characterized by jumps
whenever a large loop is disconnected. Finally, as all loops are removed, the MST ratio tends to 1.

We compared these results to the MST ratios of the largest optimized DBNs that were computationally
feasible ($N=722$ nodes) with reasonable values of the cost parameter $\gamma\in\{0.25, 0.5, 0.75\}$.
Despite the fact that the largest DBNs are smaller than real leaf networks by a factor
between approximately 20 and 50 and that the real leaf networks show considerably more variation,
the MST ratio curves are comparable,
demonstrating that
optimal DBNs exhibit similar topological features as real leaf networks [Fig.~\ref{fig:leaf-topology} (b)].
The same is not true when the nonzero $\kappa_e$ are randomly shuffled and the MST ratios recomputed
[Fig.~\ref{fig:leaf-topology} (c)], demonstrating that weighted topology and hierarchical structure
of real leaf networks are quantitatively reproduced in optimized DBNs.

\begin{figure}
    \centering
    \includegraphics[width=\textwidth]{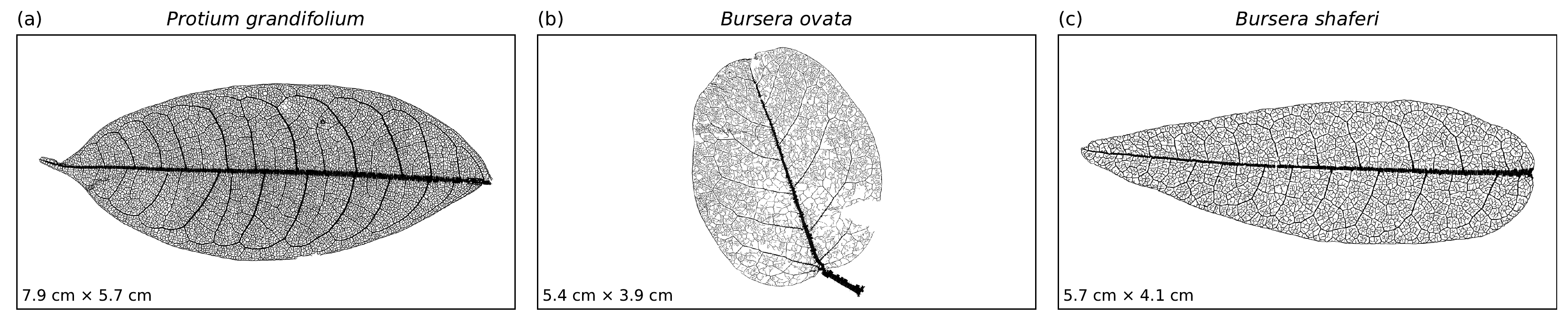}
    \caption{Discretized leaf networks used to compare to optimized DBNs.
    (a) \emph{Protium grandifolium}. $N=45324$ nodes, $E=53772$ edges.
    (b) \emph{Bursera ovata}. $N=24692$ nodes, $E=26692$ edges.
    (c) \emph{Bursera shaferi}. $N=19274$ nodes, $E=21487$ edges.
    The dimensions of each panel are indicated in the corners.}
    \label{fig:leaves}
\end{figure}

\end{document}